\documentclass[12pt,american,english]{amsart}
\usepackage[T1]{fontenc}
\usepackage{a4wide}
\usepackage{graphicx}
\usepackage{verbatim}
\usepackage{amsmath}
\usepackage{amssymb}
\usepackage{version}

 \theoremstyle{plain}
 \newtheorem{thm}{Theorem}[section]
 \numberwithin{equation}{section} 
 \numberwithin{figure}{section} 
 \theoremstyle{plain}
 \newtheorem{pro}[thm]{Proposition} 
 \newtheorem{lem}[thm]{Lemma} 
 \newtheorem{cor}[thm]{Corollary} 
\newtheorem{remark}[thm]{Remark}
\newtheorem{definition}[thm]{Definition}

\usepackage{babel}

\newcommand{\Cy}{\mathcal{C}}
\newcommand{\coloneq}{:=}
\newcommand{\lt}{<}
\newcommand{\dimostrazione}{\noindent{\bf Proof.}\phantom{X}}
\newcommand{\OpW}{{\rm Op}^{{\rm W}}_N}
\newcommand{\OpWR}{{\rm Op}^{{\rm W},\IR^2}_N}
\newcommand{\OpAW}{{\rm Op}^{{\rm AW}\!,\sigma}_N}
\newcommand{\rmd}{{\rm d}}

\newcommand{\bq}{{\bf q}}
\newcommand{\bp}{{\bf p}}

\newcommand{\bga}{\boldsymbol{\gamma}}
\newcommand{\balpha}{\boldsymbol{\alpha}}
\newcommand{\bbeta}{\boldsymbol{\beta}}
\renewcommand{\Gamma}{\varGamma}
\renewcommand{\Theta}{\varTheta}
\renewcommand{\Psi}{\varPsi}
\renewcommand{\Re}{{\mathfrak{Re}}}

\begin{document}
\newcommand{\nwc}{\newcommand}
\nwc{\nwt}{\newtheorem}


\nwc{\mf}{\mathbf} 
\nwc{\blds}{\boldsymbol} 
\nwc{\ml}{\mathcal} 


\nwc{\lam}{\lambda}
\nwc{\del}{\delta}
\nwc{\Del}{\Delta}
\nwc{\Lam}{\Lambda}


\nwc{\IA}{\mathbb{A}} 
\nwc{\IB}{\mathbb{B}} 
\nwc{\IC}{\mathbb{C}} 
\nwc{\ID}{\mathbb{D}} 
\nwc{\IE}{\mathbb{E}} 
\nwc{\IF}{\mathbb{F}} 
\nwc{\IG}{\mathbb{G}} 
\nwc{\IH}{\mathbb{H}} 
\nwc{\IN}{\mathbb{N}} 
\nwc{\IP}{\mathbb{P}} 
\nwc{\IQ}{\mathbb{Q}} 
\nwc{\IR}{\mathbb{R}} 
\nwc{\IS}{\mathbb{S}} 
\nwc{\IT}{\mathbb{T}} 
\nwc{\IZ}{\mathbb{Z}} 



\nwc{\va}{{\bf a}}
\nwc{\vb}{{\bf b}}
\nwc{\vc}{{\bf c}}
\nwc{\vd}{{\bf d}}
\nwc{\ve}{{\bf e}}
\nwc{\vf}{{\bf f}}
\nwc{\vg}{{\bf g}}
\nwc{\vh}{{\bf h}}
\nwc{\vi}{{\bf i}}
\nwc{\vj}{{\bf j}}
\nwc{\vk}{{\bf k}}
\nwc{\vl}{{\bf l}}
\nwc{\vm}{{\bf m}}
\nwc{\vn}{{\bf n}}
\nwc{\vo}{{\it o}}
\nwc{\vp}{{\bf p}}
\nwc{\vq}{{\bf q}}
\nwc{\vr}{{\bf r}}
\nwc{\vs}{{\bf s}}
\nwc{\vt}{{\bf t}}
\nwc{\vu}{{\bf u}}
\nwc{\vv}{{\bf v}}
\nwc{\vw}{{\bf w}}
\nwc{\vx}{{\bf x}}
\nwc{\vy}{{\bf y}}
\nwc{\vz}{{\bf z}}



\nwc{\bk}{\blds{k}}
\nwc{\bm}{\blds{m}}
\nwc{\bn}{\blds{n}}
\nwc{\bv}{\blds{v}}
\nwc{\bw}{\blds{w}}
\nwc{\bx}{\blds{x}}
\nwc{\bxi}{\blds{\xi}}
\nwc{\by}{\blds{y}}
\nwc{\bz}{\blds{z}}


\nwc{\cA}{\ml{A}}
\nwc{\cB}{\ml{B}}
\nwc{\cC}{\ml{C}}
\nwc{\cD}{\ml{D}}
\nwc{\cE}{\ml{E}}
\nwc{\cF}{\ml{F}}
\nwc{\cG}{\ml{G}}
\nwc{\cH}{\ml{H}}
\nwc{\cI}{\ml{I}}
\nwc{\cJ}{\ml{J}}
\nwc{\cK}{\ml{K}}
\nwc{\cL}{\ml{L}}
\nwc{\cM}{\ml{M}}
\nwc{\cN}{\ml{N}}
\nwc{\cO}{\ml{O}}
\nwc{\cP}{\ml{P}}
\nwc{\cQ}{\ml{Q}}
\nwc{\cR}{\ml{R}}
\nwc{\cS}{\ml{S}}
\nwc{\cT}{\ml{T}}
\nwc{\cU}{\ml{U}}
\nwc{\cV}{\ml{V}}
\nwc{\cW}{\ml{W}}
\nwc{\cX}{\ml{X}}
\nwc{\cY}{\ml{Y}}
\nwc{\cZ}{\ml{Z}}


\nwc{\ad}{\rm ad}
\nwc{\eps}{\epsilon}
\nwc{\ep}{\epsilon}
\nwc{\vareps}{\varepsilon}

\def\tr{{\rm tr}}
\def\Tr{{\rm Tr}}
\def\i{{\rm i}}
\def\mi{{\rm i}}
\def\e{{\rm e}}
\def\sq2{\sqrt{2}}
\def\sqn{\sqrt{N}}
\def\defi{\stackrel{\rm def}{=}}
\def\t2{{\mathbb T}^2}
\def\tc{T_{\mathbb C}}
\def\s2{{\mathbb S}^2}
\def\hn{\mathcal{H}_{N}}
\def\shbar{\sqrt{\hbar}}
\def\S{\mathcal{S}}
\def\A{\mathcal{A}}
\def\N{\mathbb{N}}
\def\T{\mathbb{T}}
\def\R{\mathbb{R}}
\def\Z{\mathbb{Z}}
\def\C{\mathbb{C}}
\def\O{\mathcal{O}}
\def\Sp{\mathcal{S}_+}

\def\Nto8{\xrightarrow{N\to \infty}}
\def\hto0{\xrightarrow{h\to 0}}
\def\htoo{\stackrel{h\to 0}{\longrightarrow}}
\def\rto0{\xrightarrow{r\to 0}}

\renewcommand{\qed}{\hfill$\blacksquare$}

\providecommand{\abs}[1]{\lvert#1\rvert}
\providecommand{\norm}[1]{\lVert#1\rVert}

\nwc{\la}{\langle}
\nwc{\ra}{\rangle}
\nwc{\lp}{\left(}
\nwc{\rp}{\right)}

\nwc{\bal}{\begin{align}}
\nwc{\bequ}{\begin{equation}}
\nwc{\ben}{\begin{equation*}}
\nwc{\bea}{\begin{eqnarray}}
\nwc{\bean}{\begin{eqnarray*}}
\nwc{\bit}{\begin{itemize}}

\nwc{\eal}{\end{align}}
\nwc{\eequ}{\end{equation}}
\nwc{\een}{\end{equation*}}
\nwc{\eea}{\end{eqnarray}}
\nwc{\eean}{\end{eqnarray*}}
\nwc{\eit}{\end{itemize}}

\title{Quantum Variance and Ergodicity for the baker's map}
\author{M. Degli Esposti,  S. Nonnenmacher \and B. Winn}
\address{Department of Mathematics,
University of Bologna Piazza di Porta S. Donato, 5 40127 Bologna,
Italy ({\tt desposti@dm.unibo.it, winn@dm.unibo.it}) 
\newline\newline 
Service de Physique
Th\'eorique, CEA/DSM/PhT Unit\'e de recherche associ\'ee au CNRS
CEA/Saclay 91191 Gif-sur-Yvette c\'edex, France ({\tt
nonnen@spht.saclay.cea.fr})\newline\newline 
Department of Mathematics ,
Texas A\&M University,
College Station,
TX 77843-3368, USA ({\tt Brian.Winn@math.tamu.edu})}

\date{22nd March 2005}

\begin{abstract}
We prove an Egorov theorem, or quantum-classical correspondence, for the
quantised baker's map, valid up to the Ehrenfest time.
This yields a logarithmic upper bound for
the decay of the quantum variance, and, as a corollary, a quantum ergodic
theorem for this map.
\end{abstract}

\maketitle


\section{Introduction}

The correspondence principle of quantum mechanics suggests that in the
classical limit the behaviour of quantum systems reproduces that of
the system's classical dynamics. It is becoming clear
that to understand this process fully represents a challenge not only
to methods of semiclassical analysis, but also the modern theory
of dynamical systems.

For a broad class of smooth Hamiltonian systems it has been proved
that if the system is ergodic, then, in the classical
limit, almost
all eigenfunctions of the corresponding quantum mechanical
Hamiltonian operator become equidistributed with respect to the
{\it natural} measure (Liouville) over the energy shell.
This is the content of the so-called {\em quantum ergodicity
theorem} \cite{Scn,Zel,CdV,HMR}.

This mathematical result, even if it can be considered quite mild
from the physical point of view, still constitutes one of the few
rigorous results concerning the properties of quantum eigenfunctions
in the classical limit, and it still leaves open the
possible existence of {\it exceptional} subsequences of eigenstates which
might converge to other invariant measures. In the last few years
a certain number of works have explored this
mathematically and physically interesting issue. While exceptional
subsequences can be present for some hyperbolic systems with
extremely high quantum degeneracies \cite{FDBN}, it is believed that they
do not exist for a {\it typical} chaotic system
(by chaotic, we generally mean that the system is ergodic and mixing).
The uniqueness of the classical limit for the quantum diagonal
matrix elements is called {\it quantum unique
ergodicity} (QUE) \cite{RS,Sar}.
There have been interesting recent results
in this direction for Hecke eigenstates of the Laplacian on
compact arithmetic surfaces \cite{Lin}, using methods which combine
rigidity properties of
semi-classical measures with purely dynamical systems theory.

The model studied in the present paper is not a Hamiltonian flow, but
rather a discrete-time symplectic map on the 2-dimensional torus phase space.
In the case of quantised hyperbolic
automorphisms of the 2-torus (``quantum cat maps''), QUE has been proven
along a subsequence of Planck's constants \cite{DEGI, KR2},
and for a certain class of eigenstates (also called ``Hecke'' eigenstates)
\cite{KR1} without restricting Planck's constant.
QUE has also been proved in the case of some uniquely ergodic
maps \cite{MR,Rosen}.
Quantum (possibly non-unique) ergodicity has been shown
for some ergodic maps which are smooth by parts, with
discontinuities on a set of zero Lebesgue measure \cite{DBDE,MOK,DEO'KW}.
Discontinuities generally produce
diffraction effects at the quantum level, which need to be taken care of
(this problem also appears in the case of Euclidean billiards with non-smooth
boundaries \cite{GL,ZZ}). Most proofs of quantum ergodicity
consist of showing
that the quantum variance defined below (equation (\ref{e:variance}))
vanishes in the classical limit.

To state our results we now turn to the specific dynamics considered in
the present article. We take as classical dynamical system the
baker's map on $\t2$, the 2-dimensional torus  \cite{AA}.
For any even positive integer $N\in 2\IN$ ($N$ is the
inverse of Planck's constant $h$), this map can be quantised into a unitary operator
(propagator) $\hat{B}_N$ acting on an $N$-dimensional Hilbert space. The 
{\it quantum
variance} measures the average equidistribution of the eigenfunctions
$\{ \varphi_{N,j}\}_{j=0}^{N-1}$ of $\hat{B}_N$:
\begin{equation}
  \label{e:variance}
  S_2(a,N)\coloneq \frac1N\sum_{j=0}^{N-1}\Big|
\la\varphi_{N,j},\OpW(a)\varphi_{N,j}\ra-\int_{\t2} a(q,p)\,\rmd q \rmd p\Big|^2\,.
\end{equation}
Here $a$ is some smooth function (observable) on $\t2$ and $\OpW(\cdot)$
is the Weyl quantisation
mapping a classical observable to a
corresponding quantum operator.
The quantised baker's map (or some variant of it) is a well-studied example
in the physics
literature on quantum chaology \cite{BV,Sa,SV1,O'CTH,Lak,Kap,Zycz},
which motivated our desire
to provide rigorous proofs for both the quantum-classical correspondence
and quantum ergodicity.

In this paper we prove a logarithmic upper bound on
the decay of the quantum variance (see theorem \ref{main} below),
which implies
quantum ergodicity as a byproduct (corollary~\ref{maincor}).
A similar upper bound
was first obtained by Zelditch \cite{Z2} in the case of the
geodesic flow of a compact negatively curved Riemannian
manifold, and was generalized by Robert \cite{Rob} to more general
ergodic Hamiltonian systems. Both are using some control on the 
the rate of classical ergodicity (Zelditch
also proved similar upper bounds for
higher moments of the matrix elements). 
The main semiclassical ingredient needed for all proofs of quantum ergodicity
is some control on the correspondence between quantum and classical
evolutions of observables, namely some \emph{Egorov estimate}. As for billiard
flows \cite{Fa}, such a correspondence can only hold for observables supported away from
the set of discontinuities.
We establish this correspondence for the
quantum baker's map in section~\ref{s:Egorov}, generalizing previous results
\cite{DBDE} for a subclass of observables (a Egorov theorem was already proven
in \cite{RubSal} for a different quantisation of the baker's map).
Some related results can be found in \cite{BGP,BR} for the case of
smooth Hamiltonian systems.
To obtain this Egorov estimate, we study the propagation of
coherent states (Gaussian wavepackets): they provide a convenient way
to ``avoid'' the set of discontinuities.
The correspondence will hold up to times of the order
of the \emph{Ehrenfest time}
\bequ\label{e:Ehrenfest}
T_{\rm E}(N)\coloneq \frac{\log N}{\log 2}
\eequ
(here $\log 2$ is the positive Lyapunov exponent of the classical baker's map).

Equipped with this estimate, one could apply the general
results of \cite{MOK} to prove that the quantum variance semiclassically
vanishes. We prefer to generalise the method of
\cite{Sch2} (applied to smooth maps or flows) 
to our discontinuous baker's map. This method, inspired by
some earlier heuristic calculations \cite{FP,Wil,EFK}, yields a
logarithmic upper bound for the variance. It relies on the decay
of classical correlations (mixing property), which is related, yet not 
equivalent, with the control on the rate of ergodicity used in 
\cite{Z2,Rob}.

Our main result is the following theorem.
\begin{thm}\label{main}
 For any observable $a\in
C^\infty(\t2)$, there is a constant $C(a)$ depending only on $a$,
such that the quantum variance over the eigenstates of $\hat B_N$
satisfies:
$$
\forall N\in 2\IN,\qquad S_2(a,N) \leq\frac{C(a)}{\log N}\,.
$$
\end{thm}
We believe that this method can be extended to
any piecewise linear map satisfying a fast mixing. We also can speculate that
the method would work for non-linear piecewise-smooth maps, although in
that case the propagation of
coherent states should be analysed in more detail (see remark \ref{remarky}).

The upper bound in theorem \ref{main} seems far from being sharp.
The heuristic calculations in \cite{FP,Wil,EFK}
suggest that the quantum variance decays
like $V(a)\,N^{-1}$ where the prefactor $V(a)$ is the
\emph{classical variance} of
the observable $a$, appearing in the central limit theorem. This has
been conjectured to be the true decay rate for a ``generic'' Anosov system.
The decay of quantum variance has been studied numerically in
\cite{EFK} for the baker's map and \cite{BSS} for Euclidean billiards; in
both cases, the results seems to be compatible with a decay $\asymp N^{-1}$;
however,
a discrepancy of around 10\% was noted between the observed
and conjectured prefactors. This was attributed to the low values of
$N$ (or energy in the case of billiards) considered. A more recent numerical
study of a chaotic billiard, at higher energies, still reveals some
(smaller) deviations from the conjectured law \cite{Bar}, leaving open the possibility
of a decay $\asymp N^{-\gamma}$ with $\gamma\neq 1$.

A decay of the form $\tilde V(a)\,N^{-1}$ (with an explicit factor $\tilde V(a)$)
could be rigorously proven for two particular Anosov
systems, using their rich arithmetic structure \cite{KR1,LS,RuSo}.
In both cases, the prefactor $\tilde V(a)$ generally
differs from the classical variance $V(a)$, which is attributed to
the arithmetic properties of the systems, which potentially makes them
``non-generic''. Algebraic decays have also been
proven for some uniquely ergodic (non-hyperbolic)
maps \cite{MR,Rosen}, by pushing the Egorov property to times of
order $\cO(N)$.

The rigorous investigation of the quantum variance thus remains an important open problem in
quantum chaology \cite{sar2}.

Quantum ergodicity follows from theorem~\ref{main} as a corollary:
\begin{cor}\label{maincor}
For each $N\in 2\IN$ there exists a subset $J_N\subset\{1,\ldots,N\}$,
with $\frac{\#J_N}{N}\Nto8 1$, such that for any $a\in
C^{\infty}(\t2)$ and any sequence $(j_N\in J_N)_{N\in 2\IN}$, 
\begin{equation}\label{e:QE}
\lim_{N\to\infty} \langle \varphi_{N,j_N}, \OpW(a)\varphi_{N,j_N}\rangle
=\int_{\t2} a(\bx) \rmd\bx\,.
\end{equation}
\end{cor}
This generalises a result of \cite{DBDE} to any
observable $a\in C^{\infty}(\t2)$ (previously only observables of the
form $a=a(q)$ could be handled). The restriction to a subset $J_N$ is the
``almost all'' clarification in quantum ergodicity.

The paper is organised as follows. In section \ref{s:classical} we briefly
describe
the classical baker's map on $\t2$. In section~\ref{s:quantum}, we recall
how this
map can be quantised \cite{BV} into an $N\times N$ unitary matrix. We then
describe the action of the  quantised baker map on coherent states
(proposition~\ref{prop:zero}). This is the first step towards the
Egorov estimates proven in section~\ref{s:EGOROV} (theorems~\ref{thm:egorov-n}
and \ref{thm:egorov}, which shows the correspondence up to the Ehrenfest time).
The first part of that section (subsection~\ref{s:W-AW})
compares the Weyl and anti-Wick quantisations,
for observables which become more singular when $N$ grows.
This technical step is necessary to obtain Egorov estimates
for times $\asymp \log N$. In the final section, we implement the
method of \cite{Sch2} to the quantum baker's map, using our Egorov
estimates up to logarithmic times, and prove theorem~\ref{main}.

\bigskip

{\bf Acknowledgments:} We are grateful to R.~Schubert for communicating to us his
results \cite{Sch2} prior to publication, and for interesting comments. 
We also thank S.~De~Bi\`evre, M.~Saraceno, N.~Anantharaman, A.~Martinez and
S.~Graffi for interesting discussions and comments.

This work has been partially supported by the European Commission
under the Research Training Network (Mathematical Aspects of
Quantum Chaos) HPRN-CT-2000-00103 of the IHP Programme.


\section{The classical baker's map\label{s:classical}}

The baker's map\footnote{The name refers to the cutting and stretching
mechanism in the dynamics of the map which is reminiscent of the procedure
for making bread.
Hence we write the word ``baker'' with a lower case ``b''.}
is the prototype model for discontinuous
hyperbolic systems, and it has been extensively studied in the
literature. Standard results may be found in \cite{AA}, while
the exponential mixing property was analyzed by \cite{Has}, and also
derives from the results of \cite{Chernov}.
Here, for the
sake of fixing notations, we restrict ourself to recalling the
very basic definitions and properties, referring the reader to the
above references for more details concerning the
ergodic properties of the map.

We identify the torus $\t2$ with the square $[0,1)\times
[0,1)$. The first (horizontal) coordinate $q$ represents the ``position'', while the
second (vertical) represents the ``momentum''. In our notations, $\bx=(q,p)$ will
always represent a phase space point, either on $\IR^2$ or on its quotient $\t2$.

The baker's map is defined as the following piecewise linear bijective
transformation on $\t2$:
\bequ\label{e:clas.map}
B(q,p)=(q',p')=\begin{cases} (2q,p/2),&\text{if}\ q\in[0,1/2),\\
(2q-1,(p+1)/2),&\text{if}\ q\in[1/2,1).\end{cases}
\eequ
The
transformation is discontinuous on the following subset of $\t2$:
\bequ \cS_1\coloneq\{p=0\}\cup\{q=0\}\cup\{q=1/2\}, \eequ and smooth
everywhere else. If we consider iterates of the map, the
discontinuity set becomes larger: for any $n\in\IN$, the map $B^n$
is piecewise linear, and discontinuous on the set
$$
\cS_n\coloneq \{p=0\}\cup\bigcup_{j=0}^{2^n-1}
\left\{q=\frac{j}{2^n}\right\}\,,
$$
while its inverse $B^{-n}$ is discontinuous on the set $\cS_{-n}$ obtained from
$\cS_n$ by exchanging the $q$ and $p$ coordinates.
Clearly, the discontinuity set becomes dense in $\t2$ as $|n|\to\infty$.
The map is area preserving and uniformly hyperbolic outside the
discontinuity set, with constant Lyapunov exponents $\pm \log 2$ and positive
topological entropy (see below). The stable (resp.\ unstable) manifold is made of
vertical (resp.\ horizontal) segments.

A nice feature of this map lies in a simple symbolic coding for its
orbits. Each real number $q\in [0,1)$ can be associated with a
binary expansion
$$
q= \cdot\,\ep_0\ep_1\ep_2\ldots\qquad (\ep_i\in\{0,1\}).
$$
This representation is one-to-one if we forbid expansions
of the form $\cdot\ep_0\ep_1\ldots 111\ldots$
Using the same representation for the $p$-coordinate:
$$
p=\cdot\,\ep_{-1}\ep_{-2}\ldots,
$$
a point $\bx=(q,p)\in\t2$ can be represented by the doubly-infinite sequence
$$
\bx=\ldots \ep_{-2}\ep_{-1}\cdot\ep_0\ep_1\ldots
$$
Then, one can easily check that the baker's map acts on this representation
as a symbolic shift:
\bequ\label{e:symbolic}
B(\ldots\ep_{-2}\ep_{-1}\cdot\ep_0\ep_1\ldots)=\ldots\ep_{-2}\ep_{-1}\ep_0\cdot\ep_1\ldots
\eequ
{}From this symbolic representation, one gets the Kolmogorov-Sinai entropy
of the map, $h_{\rm KS}=\log 2$, as well as exponential mixing properties \cite{Chernov,Has}:
there exists $\Gamma>0$ and $C>0$ such that,
for any smooth observables $a,\,b$ on $\t2$, the correlation function
\bequ
{\mathcal K}_{ab}(n)\coloneq\int_{\t2} a(\bx)\,b(B^{-n}\bx)\,\rmd
\bx- \int_{\t2} a(\bx)\,\rmd \bx \int_{\t2} b(\bx)\,\rmd \bx
\eequ
is bounded as
\bequ\label{e:mixing}
|{\mathcal K}_{ab}(n)|\leq C\,\norm{a}_{C^1}\,\norm{b}_{C^1}\,
\e^{-\Gamma|n|}\,.
\eequ
According to \cite{Has}, one can take for $\Gamma$ any number smaller than $\log 2$.


\section{Quantised baker's map\label{s:quantum}}
The quantisation of the 2-torus phase space is now
well-known and we refer the reader to \cite{DEG}, here describing only the
important facts. The quantisation of an area-preserving map on the torus
is less straightforward, and in general it contains some arbitrariness.
The quantisation of linear symplectomorphisms of the 2-torus (or ``generalised
Arnold cat maps'')
was first considered
in \cite{HB}, and the case of nonlinear perturbations of cat maps was
treated in \cite{BasOz} (quantum ergodicity was proven for these maps in \cite{BDeB}). 
The scheme we present below, specific for the
baker's map, was introduced in \cite{BV}.

We start by defining the quantum Hilbert space associated to the torus phase space.
For any $\hbar\in (0,1]$, we consider the quantum translations (elements of the Heisenberg group)
$\hat T_{\bv}=\e^{\i(v_2\hat{q}-v_1\hat{p})/\hbar}$, $\bv\in\IR^2$, acting on $L^2(\IR)$
 and by extension
on $\cS'(\IR)$.  We then define the space of distributions
$$
\cH_\hbar=\{\psi\in\cS'(\IR),\ \hat T_{(1,0)}\psi=\hat T_{(0,1)}\psi=\psi\}\,.
$$
These are distributions $\psi(q)$ which are $\IZ$-periodic, and such that their
$\hbar$-Fourier transform
\bequ\label{e:hbar-FT}
(\hat F_\hbar \psi)(p)\coloneq \int_{-\infty}^\infty \psi(q)\,
\e^{-\i qp/\hbar}\,\frac{\rmd q}{\sqrt{2\pi\hbar}}
\eequ
is also $\IZ$-periodic.

One easily shows that this space is nontrivial iff $(2\pi\hbar)^{-1}=N\in\IN$,
which we will always assume from now on.
This space can be obtained as the image of $L^2(\IR)$ through the ``projector''
\bequ\label{e:projector}
\hat P_{\t2} = \sum_{\bm\in\IZ^2} (-1)^{N m_1 m_2}\,\hat T_{\bm}
=\Big(\sum_{m_2\in\IZ}\hat T_{0,m_2}\Big)\;\Big(\sum_{m_1\in\IZ}\hat T_{m_1,0}\Big)\,.
\eequ
$\cH_\hbar=\hn$ then forms an $N$-dimensional vector space of distributions, admitting a
``position representation''
\bequ\label{e:q-basis}
\psi(q)=\frac{1}{\sqrt{N}}\sum_{j=0}^{N-1}\sum_{\nu\in\IZ}\psi_j\
\delta\!\left(q-\frac{j}{N}+\nu\right)=:\sum_{j=0}^{N-1}\psi_j\,\bq_j(q),
\eequ
where each coefficient $\psi_j\in\IC$. Here we have denoted by
$\{\bq_j\}_{j=0}^{N-1}$ the canonical (``position'') basis for $\cH_N$.

This space can be naturally equipped with
the Hermitian inner product:
\begin{equation}
  \label{e:innerprod}
\la \bq_j,\bq_k\ra=\delta_{jk}\Longrightarrow
\langle \psi,\omega\rangle \coloneq \sum_{j=0}^{N-1} \overline{\psi_j}\,
\omega_j\,.
\end{equation}
Since $\hn$ is the image of $\cS(\IR)$ through the ``projector'' \eqref{e:projector},
any state $\psi\in\cH_N$ can be constructed by projecting some Schwartz function $\Psi(q)$.
The decomposition on the RHS of \eqref{e:projector} suggests that we may first periodicise
in the $q$-direction, obtaining a periodic function $\Psi_{\Cy}(q)$; such a wavefunction
describes a state living
in the \emph{cylinder} phase space $\Cy=\IT\times\IR$.
The torus state $\psi(q)$
is finally obtained by periodicising $\Psi_{\Cy}$ in the Fourier variable; equivalently,
the $N$ components of $\psi$ in the
basis $\{\bq_j\}$ are obtained by sampling this
function at the points $q_j=\frac{j}{N}$:
\begin{equation}
  \psi_j= \frac{1}{\sqrt{N}}\,\Psi_{\Cy}\Big( \frac{j}{N}\Big)\,, \qquad 0\leq j \lt N.
\end{equation}
The $\hbar$-Fourier transform $\hat F_\hbar$ (seen as a linear operator on $\cS'(\IR)$)
leaves the space $\hn$ invariant. On the basis $\{\bq_j\}$, it acts as an $N\times N$
unitary matrix $\hat F_N$
called the ``discrete Fourier transform'':
\bequ\label{e:FT}
(\hat F_N)_{kj}=
 \frac{1}{\sqrt{N}}\;\e^{-2\i\pi kj/N}\,,\quad k,j=0,\ldots,N-1\,.
\eequ
$\hat F_\hbar$ quantises the rotation by $-\pi/2$ around the origin, $F\,(q_0,p_0)= (p_0,-q_0)$.
As a result, $\hat F_N$
maps the ``position basis'' $\{\bq_j\}$ onto the ``momentum basis''
$\{\bp_j\}$:
$$
\bp_j=\sum_{k=0}^{N-1} (\hat F^{-1}_N)_{kj}\,\bq_k\,.
$$
The quantised baker's map $\hat{B}_N$ was introduced by Balazs and Voros \cite{BV}.
They require $N$ to be an \emph{even} integer, and prescribe the following matrix
in the basis $\{\bq_j\}$:
\begin{equation}
  \label{e:baker}
  \hat{B}_N \coloneq
(\hat{F}_N)^{-1} \hat{B}_{N,{\rm mix}}\,,\qquad\text{with}\qquad
\hat B_{N,{\rm mix}}\coloneq
\begin{pmatrix} \hat F_{N/2}&0\\0&\hat F_{N/2}\end{pmatrix}\,.
\end{equation}

This definition was slightly modified by Saraceno \cite{Sa}, in order to restore
the parity symmetry of the classical map. Although we will concentrate on
the map (\ref{e:baker}), all our results also apply to this modified
setting.


\subsection{Notations\label{s:notations}}$ $

Since we will be dealing with quantities depending on Planck's constant $N$ (plus
possibly other parameters), all asymptotic notations will refer to the
classical limit $N\to\infty$.

The notations $A=\cO(B)$ and $A\ll B$ both mean that there exists a
constant $c$ such that for any $N\geq 1$, $|A(N)|\leq c|B(N)|$.
Writing $A=\cO_r(B)$ and $A\ll_r B$ means that the constant $c$ depends on
the parameter $r$.
Similarly $A={\rm o}(B)$ and $A<<B$ both mean that $\lim_{N\to\infty}
\frac{A(N)}{B(N)}=0$.
By $A\asymp B$ we mean that $A\ll B$ and $B\ll A$ simultaneously.
We indicate by $A\sim B$
the more precise asymptotics $\lim_{N\to\infty}\frac{A(N)}{B(N)}=1$.

We use the convention for number sets that $\IN:= \{1,2,3,\ldots\}$ and
$\IN_0:= \IN \cup \{0\}$. Also $\IR_{+}:=[0,\infty)$, as usual.

We will use various norms. We denote by $\norm{\cdot}_{\cH_N}$ the
norm on $\cH_N$ defined as $\norm{\psi}_{\cH_N}^2=\langle\psi,\psi\rangle$.
Unless stated otherwise, $\norm{\cdot}$ will
refer to the norm on bounded operators on $\cH_N$, also denoted by
$\norm{\cdot}_{\cB(\hn)}$.
The Hilbert-Schmidt scalar product of two operators $A,\,B$ on $\hn$ will be denoted by
\bequ\label{e:HS}
\Big\la A,B\Big\ra \coloneq \frac{1}{N}\ \Tr(A^\dagger B)\,.
\eequ

Other norms describe classical observables (smooth functions $f$ on $\t2$).
The sup-norm will be denoted by $\norm{f}_{C^0}$,
and for any $j>0$, the $C^j$-norm is defined as
$$
\norm{f}_{C^j}\coloneq \sum_{0\leq|\bga|\leq j} \norm{\partial^{\bga}f}_{C^0}\,.
$$
Here $\bga=(\gamma_1,\gamma_2)\in\IN^2_0$ denotes the multiindex of differentiation:
$\partial^{\bga}=\partial^{\gamma_1}_q\,\partial^{\gamma_2}_p$, and
$|\bga|\coloneq\gamma_1+\gamma_2$.

Because we want to consider large time evolution, namely times $n\asymp \log N$,
we need to
consider (smooth) functions which depend on Planck's constant $1/N$.
Indeed, starting from a given smooth function $a$,
its evolution $a\circ B^{-n}$ fluctuates more and more strongly along the vertical direction,
while it is smoother and smoother along the horizontal one as $n\to\infty$
(assuming $a$ is supported away from the discontinuity set $\cS_n$).
For this reason, we introduce the following spaces of functions \cite[chapter~7]{DS}:
\begin{definition}\label{d:S_alpha}
For any $\balpha=(\alpha_1,\alpha_2)\in\IR_+^2$,
we call $S_{\balpha}(\t2)$ the space of $N$-dependent smooth functions
$f=f(\cdot,N)$ such that, for any multiindex ${\bga}\in \IN^2_0$,
the quantity
$$
C_{\balpha,\bga}(f)\coloneq\sup_{N\in\IN}
\frac{\norm{\partial^{\bga} f(\cdot,N)}_{C^0}}{N^{\balpha\cdot\bga}}
$$
is finite (here $\balpha\cdot\bga=\alpha_1\gamma_1+\alpha_2\gamma_2$).
The seminorms
$C_{\balpha,\bga}$ ($\bga\in\IN^2_0$) endow $S_{\balpha}(\t2)$ with the structure of a Fr\'echet space.
\end{definition}


\section{Coherent states on $\t2$\label{s:CS}}

Our proof of the quantum-classical correspondence will use coherent states
on $\t2$. Below we define them, and collect some useful
properties. More comprehensive details and proofs
may be found in \cite{Fo,perelomov,LebVor,BDeB,BonDB}.

We define a (plane) coherent state at the origin with squeezing
$\sigma>0$ through its wavefunction $\Psi_{0,\sigma}\in\cS(\IR)$ (we will always omit
indicate the $\hbar$-dependence):
\begin{equation}
  \label{e:planeCS}
  \Psi_{0,\sigma}(q)\coloneq \left( \frac{\sigma}{\pi\hbar}\right)^{1/4}
  \e^{-\frac{\sigma q^2}{2\hbar}}.
\end{equation}
The (plane) coherent state at the point $\bx=(q_0,p_0)\in\IR^2$ is obtained by
applying a quantum translation $\hat{T}_{\bx}$ to the state above, which yields:
\begin{align*}
  \Psi_{\bx,\sigma}(q)&\coloneq  \left( \frac{\sigma}{\pi\hbar}\right)^{1/4}
   \,\e^{-\i \frac{p_0 q_0}{2\hbar}}\,
\e^{\i \frac{p_0 q}{\hbar}}\,
\e^{\frac{-\sigma (q-q_0)^2}{2\hbar}} \\
&= (2N\sigma)^{1/4}\e^{-\pi\i N q_0p_0 + 2\pi\i N p_0 q -\sigma N \pi
 (q-q_0)^2}.
\end{align*}
(In the second line, we took $\hbar=(2\pi N)^{-1}$, as is required if we want to project on the
torus). From here we obtain a coherent state on the cylinder by periodicising
along the $q$-axis:
\begin{equation}\label{e:cylCS}
  \Psi_{\bx,\sigma,\Cy}(q) \coloneq \sum_{\nu\in\IZ}
   \Psi_{\bx,\sigma}(q+\nu)\,.
\end{equation}
Finally, the coherent state on the torus is obtained by further periodicising in the Fourier
variable, or equivalently by sampling this cylinder wavefunction: its coefficients in
the canonical basis read
\bequ\label{e:torusCS}
\big(\psi_{\bx,\sigma,\t2}\big)_j=\frac{1}{\sqrt{N}}\,\Psi_{\bx,\sigma,\Cy}(j/N),\quad
j=0,\ldots,N-1.
\eequ
One can check that
$\psi_{\bx+\bm,\sigma,\t2}\propto \psi_{\bx,\sigma,\t2}$ for any $\bm\in\IZ^2$:
up to a phase, the state $\psi_{\bx,\sigma,\t2}$ depends on the projection
on $\t2$ of the point $\bx$.

In the classical limit, it will often be useful to approximate a
torus (or cylinder) coherent state by the
corresponding planar one:
\begin{lem} \label{lem:uno}
Let $q_0\in(\delta,1-\delta)$ for some $0\lt\delta\lt 1/2$. Then in the
classical limit:
\begin{equation}
  \label{e:approx}
\forall q\in[0,1),\qquad  \Psi_{\bx,\sigma,\Cy}(q) =
  \Psi_{\bx,\sigma}(q) + \cO\big((
\sigma N)^{1/4}\e^{-\pi N \sigma \delta^2}\big)\,.
\end{equation}
The error estimate is uniform for $\sigma N\geq 1$.
\end{lem}
\dimostrazione
Extracting the $\nu=0$ term in \eqref{e:cylCS}, one gets
\begin{equation*}
\forall q\in[0,1),\qquad  \Psi_{\bx,\sigma,\Cy}(q)= \Psi_{\bx,\sigma}(q)
    + \cO\Big((\sigma N)^{1/4}\,
    \e^{-\pi\sigma N \min\{|q-q_0+\nu|^2:\nu\neq 0 \}} \Big).
\end{equation*}
Now, if $q_0\in(\delta,1-\delta)$, one has $|q-q_0|\leq 1-\del$, so that
$$
\forall \nu\neq 0,\qquad  |q-q_0-\nu| \geq |\nu|-|q-q_0| \geq 1-|q-q_0| \geq\del\,.
$$
\qed

The next lemma describes how a torus coherent state
transforms under the application of the discrete Fourier transform.
\begin{lem} \label{lem:due}
For any $\bx=(q_0,p_0)\in\IR^2$,
let $F\,\bx\coloneq (p_0,-q_0)$ denote its rotation by $-\pi/2$ around the origin.
Then
\begin{equation}
\forall N\geq 1,\ \forall\sigma>0,\qquad
\hat{F}_N\,\psi_{\bx,\sigma,\t2} = \psi_{F\bx,1/\sigma,\t2}.
\end{equation}
\end{lem}
\dimostrazione
The plane coherent states, which are Gaussian wavefunctions,
are obviously covariant through
the Fourier transform $\hat F_\hbar$: a straightforward computation shows that
$$
\forall\bx\in\IR^2,\quad \hat{F}_\hbar\,\psi_{\bx,\sigma} = \psi_{F\bx,1/\sigma}\,.
$$
When $(2\pi\hbar)=N^{-1}$, we apply the projector \eqref{e:projector} to both sides of this
inequality, and remember that $\hat{F}_\hbar$ acts on $\hn$ as the
matrix $\hat{F}_N$: this means $\hat P_{\t2}\,\hat{F}_\hbar=\hat{F}_N\,\hat P_{\t2}$,
so the above covariance is carried over to the torus coherent states.
\qed


\subsection{Action of $\hat{B}_{N}$ on coherent states\label{s:action}}$ $

We assume $N$ to be an even integer, and
apply the matrix $\hat B_N$ to the coherent state $\psi_{\bx,\sigma,\t2}$, seen as
an $N$-component vector in the basis $\{\bq_j\}$.
We get nice results if the point $\bx=(q_0,p_0)$ is ``far enough''
from the singularity set $\cS_1$ (in this case $B\bx$ is well-defined).
More precisely, we define the following subsets
of $\t2$:
\begin{definition}
For any $0<\del<1/4$ and $0<\gamma<1/2$, let
\bequ\label{e:D1}
\cD_{1,\del,\gamma}\coloneq \left\{(q,p)\in\t2,\
q\in (\del,1/2-\del)\cup(1/2+\del,1-\del),\
p\in(\gamma,1-\gamma)\right\}\,.
\eequ
\end{definition}

The evolution of coherent states will be simple for states localised
in this set.

\begin{pro} \label{prop:zero}
For some parameters $\del,\,\gamma$ (which may depend on $N$),
we consider points $\bx=(q_0,p_0)\in\T^2$ in the set 
$\cD_{1,\del,\gamma}$. 
We associate to these points the phase
\begin{equation}
  \label{e:Theta}
  \Theta(\bx) = \begin{cases}
    0, & \mbox{if $q_0\in(\delta,1/2-\delta),$} \\
    \displaystyle
    q_0+\frac{p_0+1}2, & \mbox{if $q_0\in(1/2+\delta,1-\delta)$.}
    \end{cases}
\end{equation}
We assume that the squeezing $\sigma$ may also depend on $N$, remaining
in the interval $\sigma\in [1/N,N]$. From $\delta$, $\gamma$, $\sigma$ we form the
parameter
\bequ\label{e:theta}
\theta=\theta(\delta,\gamma,\sigma)\coloneq\min(\sigma\del^2,\gamma^2/\sigma)\,.
\eequ
Then, in the semiclassical limit, the coherent state $\psi_{\bx,\sigma,\t2}$ evolves
almost covariantly through the quantum baker's map:
 \bequ\label{e:evol-0}
\norm{\hat B_N\,\psi_{\bx,\sigma,\t2}-\e^{\i\pi\Theta(\bx)}\,\psi_{B\bx,\sigma/4,\t2}}_{\hn}
=\cO(N^{3/4}\sigma^{1/4}\,\e^{-\pi N\theta})\,.
\eequ
The implied constant is uniform with respect to $\delta$, $\gamma$,
$\sigma\in [1/N,N]$, and the point $\bx\in \cD_{1,\del,\gamma}$.
\end{pro}

We notice that the exponential in the above remainder will be small only if $\theta>> 1/N$,
which requires both $\sigma>>1/N$ and $\sigma<<N$. In further applications we
will always consider squeezings satisfying these properties in the limit $N\to\infty$. 

\begin{remark}
If we extend to the
full plane each of the maps given by the
two lines of equation~\eqref{e:clas.map}, we get two linear symplectic
transformation $S_0$, $S_1$,
which can be quantised on $L^2(\IR)$ by the
metaplectic transformations
$$
\hat{S}_{0,\hbar}=\hat{D}_2,\qquad
\hat{S}_{1,\hbar}= \hat{T}_{(-1,0)}\circ\hat{D}_2 \circ\hat{T}_{(0,1)}
$$
(here $[\hat D_2\,\psi](q)=2^{-1/2}\,\psi(q/2)$ is the unitary dilation by a factor $2$).
Such metaplectic transformations act covariantly on plane coherent states:
$$
\forall \sigma>0,\quad\forall \bx=(q_0,p_0)\in\IR^2,\qquad
\begin{cases}
\hat {S}_{0,\hbar}\, \Psi_{\bx,\sigma}
=\Psi_{S_0\bx,\sigma/4}\,,\\
\hat {S}_{1,\hbar}\, \Psi_{\bx,\sigma}
=\e^{\frac{\i}{2\hbar} (q_0+\frac{p_0+1}2)}\,\Psi_{S_1\bx,\sigma/4}\,.
\end{cases}
$$
The approximate covariance stated in proposition~\ref{prop:zero} is therefore
a microlocal version of this exact global covariance.
\end{remark}
\begin{remark} \label{remarky}
The fact that the
error is exponentially small is due to the piecewise-linear character of the
map $B$. Indeed, for a nonlinear area-preserving map $M$ on $\t2$, coherent states are
also transformed covariantly through $\hat M_N$, but the error term is in general
of order $\cO(N\,\Delta\bx^3)$, where $\Delta\bx$ is the ``maximal width'' of the coherent
state (here $\Del\bx=\max(\sigma,\sigma^{-1})\,N^{-1/2}$)  \cite{Sch}. Moreover,
in general
the squeezing $\sigma$ takes values in the complex half-plane $\{\Re(\sigma)>0\}$: the reason
why we can here restrict ourselves to the positive real line is due to the orientation
of the baker's dynamics.
\end{remark}

\noindent{\bf Proof of proposition~\ref{prop:zero}.}\phantom{X}

Since we already know that $\hat F_N$ acts covariantly on coherent states,
we only need to analyse the action of $\hat{B}_{N,{\rm mix}}$ (Eq.~\eqref{e:baker}).

We first consider a coherent state in the ``left'' strip
$(\del,1/2-\del)\times (\gamma,1-\gamma)$ of $\cD_{1,\del,\gamma}$.
In this case, the ``relevant''
coefficients of $\hat{B}_{N,{\rm mix}}\, \psi_{\bx,\sigma,\t2}$
are in the interval $0\leq m < \frac{N}2$:
\begin{equation}
\label{e:action:uno}
\left( \hat{B}_{N,{\rm mix}}\, \psi_{\bx,\sigma,\t2}\right)_m
 =\frac1{\sqrt{N}}\sum_{j=0}^{N/2-1} (\hat{F}_{N/2})_{mj}
 \ \Psi_{\bx,\sigma,\Cy}\!\left(\frac{j}N\right)\,.
\end{equation}
{}From the formula~\eqref{e:FT}, we have for all $0\leq j,m<N/2$:
$$
(\hat{F}_{N/2})_{mj}=\sqrt{2}\,(\hat{F}_{N})_{2m\,j}\,.
$$
Since $q_0\in(\delta,1/2-\delta)$, for any $N/2\leq j$ one
has $j/N-q_0\geq\delta$; using lemma~\ref{lem:uno}, we obtain
\begin{equation}\label{e:action:due}
\forall j\in\{ N/2,\ldots, N-1\},\qquad \Psi_{\bx,\sigma,\Cy}\!\left(
\frac{j}N\right)
= \cO\big((\sigma N)^{1/4}\,\e^{-\pi N\sigma\delta^2}\big)\,.
\end{equation}
We can therefore extend the range of summation in \eqref{e:action:uno}
to $j\in\{0,\ldots, N-1\}$,  incurring only an exponentially small error:
\begin{align}
\left( \hat{B}_{N,{\rm mix}}\; \psi_{\bx,\sigma,\t2}\right)_m
 &=\sqrt{2}\; \sum_{j=0}^{N-1} (\hat{F}_{N})_{2m\,j}\,
  \Big(\psi_{\bx,\sigma,\t2}\Big)_j + \cO( (\sigma N)^{1/4}
 \e^{-\pi N\sigma\delta^2})  \nonumber \\
 \label{e:action:quattro}
 &=\sqrt{2}\;\Big(\psi_{F\bx,1/\sigma,\t2}\Big)_{2m} +
\cO( (\sigma N)^{1/4}  \e^{-\pi N\sigma \delta^2})\,.
\end{align}
In the last step, we have used the covariance property of lemma~\ref{lem:due}.

Since $p_0\in(\gamma,1-\gamma)$ and $N/\sigma\geq 1$, 
it follows from lemma \ref{lem:uno} and simple manipulations
of plane coherent states that
\begin{align*}
 \forall q\in [0,1/2),\qquad
\sqrt{2}\;\Psi_{F\bx,1/\sigma,\Cy}(2q) &= \sqrt{2}\;\Psi_{F\bx,1/\sigma}(2q)
 + \cO\big( (N/\sigma)^{1/4}\e^{-\pi N \gamma^2/\sigma}\big)\\
 &=\Psi_{(p_0/2,-2q_0),4/\sigma}(q)
 + \cO\big( (N/\sigma)^{1/4}\e^{-\pi N \gamma^2/\sigma}\big)\\
 &=\Psi_{(p_0/2,-2q_0),4/\sigma,\Cy}(q)
+    \cO\big((N/\sigma)^{1/4}\,\e^{-\pi N \gamma^2/\sigma}\big)\,.
\end{align*}
The identity $(p_0/2,-2q_0)=FB\bx$ (valid for $\bx$ in the left strip)
inserted in \eqref{e:action:quattro} yields:
\begin{equation}\label{e:action:cinque}
\forall m\in\{0,\ldots,N/2-1\},\quad
\left( \hat{B}_{N,{\rm mix}}\; \psi_{\bx,\sigma,\t2}\right)_m
= \left(\psi_{FB\bx,4/\sigma,\t2}\right)_m
+ \cO((\sigma N)^{1/4} \e^{-\pi N \theta})
\end{equation}
($\theta$ is defined in \eqref{e:theta}, and we used the assumption $\sigma N>1$
to simplify the remainder).

The remaining coefficients $N/2\leq m \lt N$ are bounded using \eqref{e:action:due}:
\bequ
  \left( \hat{B}_{N,{\rm mix}}\; \psi_{\bx,\sigma,\t2}\right)_m
 =\frac{1}{\sqrt{N}}\;\sum_{j=N/2}^{N-1} (\hat{F}_{N/2})_{m\,j}\;
  \Psi_{\bx,\sigma,\Cy}\!\left(\frac{j}N\right)
 \label{e:action:tre}
 =\cO( (\sigma N)^{1/4} \e^{-\pi\sigma N \delta^2})\,.
\eequ
On the other hand, lemma~\ref{lem:uno} shows that
the coefficients $\left(\psi_{FB\bx,4/\sigma,\t2}\right)_m$ for $N/2\leq m\lt N$
are bounded from above by the same RHS. Hence,
equation~\eqref{e:action:cinque}
holds for all $m=0,\ldots,N-1$.
A norm estimate is obtained by multiplying this component-wise estimate
by a factor $\sqrt{N}$.

We now
apply the inverse Fourier transform and lemma~\ref{lem:due},
to obtain the part of the
theorem dealing with coherent states in the left strip of $D_{1,\del,\gamma}$.

\bigskip

A similar computation treats the case of coherent states in the right strip
of $D_{1,\del,\gamma}$. The
large components of $\psi_{\bx,\sigma,\t2}$
are in the interval $j\geq N/2$, so the second block of $\hat B_{N,{\rm mix}}$
is relevant. The analogue to \eqref{e:action:cinque}
reads, for $m\in\{N/2,\ldots,N-1\}$:
\begin{equation}
  \label{e:action:sei}
   \left( \hat{B}_{N,{\rm mix}}\; \psi_{\bx,\sigma,\t2}\right)_m
= \sqrt{\frac2N}\;\Psi_{F\bx,1/\sigma}\!\left(\frac{2m}{N}-1
\right) + \cO((\sigma N)^{1/4} \e^{-\pi N \theta})\,.
\end{equation}
Proceeding as before, we identify
\begin{align}
\forall q\in[1/2,1),\quad  \sqrt{2}\;\Psi_{F\bx,1/\sigma}(2q-1)
  &=\e^{\pi\i N(q_0+\frac{p_0+1}2)}\ \Psi_{((p_0+1)/2,-(2 q_0-1)),4/\sigma}(q) \nonumber \\
  &=\e^{\pi\i N(q_0+\frac{p_0+1}2)}\ \Psi_{FB\bx,4/\sigma,\Cy}(q)
 +  \cO((N/\sigma)^{1/4}\,\e^{-\pi N \gamma^2/\sigma})\,.
\end{align}
Applying the inverse Fourier transform we obtain the second part of the theorem.
\qed


\section{Egorov property\label{s:EGOROV}}

Our objective in this section is to control the evolution of
quantum observables through $\hat B_N$,
in terms of the corresponding classical evolution.
Namely, we want to prove an Egorov theorem of the type
\bequ\label{e:egor}
\norm{\hat B^{n}_N\,{\rm Op}_N(a)\,\hat B^{-n}_N-{\rm Op}_N(a\circ B^{-n})}
\Nto8 0.
\eequ
Here ${\rm Op}_N(a)$ is some quantisation of an observable $a\in C^\infty(\t2)$.
As explained in the introduction, to avoid the diffraction problems due to the
discontinuities of $B$, we will require
the function $a$ to be supported away from the set $\cS_n$ of discontinuities of $B^n$.
Otherwise, $a\circ B^{-n}$ may be
discontinuous, and already its quantisation poses some problems.

An Egorov theorem has been proven in \cite{RubSal} for a different
quantisation of the baker's map, also using coherent states. In
\cite[corollary~17]{DBDE} an Egorov theorem was obtained for $\hat B_N$,
but valid only
for observables of the form $a(q)$ (or $a(p)$, depending on the direction of time) and
restricting the observables to a ``good'' subspace of $\hn$ of dimension $N-{\rm o}(N)$.

Since we control the evolution of coherent states through $\hat B_N$
(proposition~\ref{prop:zero}), it is natural to use a quantisation defined
in terms of coherent states, namely the \emph{anti-Wick quantisation}
 \cite{perelomov} (see definition~\ref{d:AW} below).
However, because the quasi-covariance \eqref{e:evol-0} connects a squeezing $\sigma$ to
a squeezing $\sigma/4$, it will
be necessary to relate the corresponding quantisations $\OpAW$ and
${\rm Op}^{{\rm AW}\!,\sigma/4}_N$ to one another.
This will be done in the next subsection, by using
the \emph{Weyl quantisation} as a reference.

Besides, we want
to control the correspondence \eqref{e:egor} uniformly with respect to the
time $n$. We already
noticed that for $n>>1$, an
observable $a$ supported away from $\cS_n$ needs to fluctuate quite strongly along
the $q$-direction, while
its dependence in the $p$ variable may remain mild. Likewise,
$a\circ B^{-n}$, supported away from $\cS_{-n}$,
will strongly fluctuate along the $p$-direction.

All results in this section will be stated for two classes of
observables:
\begin{itemize}
\item general functions $f\in C^\infty(\t2)$, without any indication on
how $f$ depends on $N$. This yields
a Egorov theorem valid for time $|n|\leq (\frac{1}{6}-\eps)\,T_{\rm E}$ (with $\eps>0$ fixed), 
which will suffice to prove theorem~\ref{main} ($T_{\rm E}=T_{\rm E}(N)$ is the Ehrenfest
time~\eqref{e:Ehrenfest}).
\item functions $f\in S_{\balpha}(\t2)$ for some $\balpha\in \IR_+^2$ with
$|\balpha|<1$ (see the definition~\ref{d:S_alpha}).
Here we use more sophisticated methods in order to push the Egorov
theorem up to the times $|n|\leq (1-\epsilon) T_{\rm E}$.
\end{itemize}


\subsection{Weyl vs. anti-Wick quantizations on $\t2$\label{s:W-AW}}$ $

In this subsection,
we define and compare the Weyl and anti-Wick quantisations on the torus.
The main result is proposition~\ref{prop:weyl-AW}, which
precisely estimates the discrepancies between these quantisations, in the
classical limit. We start by recalling the definition of the Weyl quantisation
on the torus \cite{BDeB,DEG}.
\begin{definition}\label{d:Weyl}
Any function $f\in C^\infty(\t2)$ can be Fourier expanded as follows:
$$
f=\sum_{\bk\in\IZ^2} \tilde{f}(\bk)\;e_{\bk}\,,\quad\text{where}
\quad e_{\bk}(\bx)\coloneq \e^{2\pi\i \bx\wedge\bk}=\e^{2\pi\i (qk_2-pk_1)}\,.
$$
The Weyl quantisation of this function is the following operator:
\bequ\label{e:Weyl} \OpW(f)\coloneq\sum_{\bk \in
\IZ^{2}}\tilde{f}(\bk)\; T(\bk)\,,\quad\text{where}\quad T(\bk)\coloneq \hat{T}_{h\bk}\,.
\eequ
We use the same notations for translation operators $T(\bk)$
acting on either $\hn$ or $L^2(\IR)$;
in the latter case, the Weyl-quantised operator will be denoted by $\OpWR(f)$.
\end{definition}
The operators $\{T(\bk)\,;\bk\in\IZ^2\}$
acting on $L^2(\IR)$ form an independent set of of unitary operators.
On the other hand, on $\hn$ these operators satisfy
$T(\bk+N\bm)=(-1)^{\bk\wedge\bm}\,T(\bk)$. Hence, defining $\IZ_N\coloneq \{-N/2,\ldots,N/2-1\}$,
the set $\{T(\bk),\ \bk\in\IZ_N^2\}$ forms a basis of the space of operators
on $\hn$. This basis is orthonormal with respect to the Hilbert-Schmidt scalar product \eqref{e:HS}.

The Weyl quantisations on $L^2(\IR)$ and $\hn$ satisfy the following inequality
\cite[lemma~3.9]{BDeB}:
\bequ\label{e:plane-torus}
\forall f\in C^\infty(\t2),\quad\forall N\in\IN,\qquad
\norm{\OpW(f)}_{\cB(\hn)}\leq \norm{\OpWR(f)}_{\cB(L^2(\IR))}\,.
\eequ
This will allow us to use results pertaining to the Weyl quantisation
of bounded functions on the plane (see the proof of lemma~\ref{lem:CV}).

We now define a family of anti-Wick quantisations.
\begin{definition}\label{d:AW}
For any squeezing $\sigma>0$, the anti-Wick
quantisation of a function $f\in L^1(\t2)$ is the operator
$\OpAW(f)$ on $\hn$ defined as:
\bequ
\label{e:AW}
\forall\phi,\,\phi'\in\hn,\qquad
\la \phi,\OpAW(f)\,\phi'\ra \coloneq
N \int_{\t2} f(\bx)\, \la\phi,\psi_{\bx,\sigma,\t2}\ra\,
\la \psi_{\bx,\sigma,\t2},\phi'\ra\,\rmd\bx.
\eequ
\end{definition}
Both Weyl and anti-Wick quantisations map a real observable onto a Hermitian operator.
As opposed to the Weyl quantisation, the anti-Wick quantisation enjoys the important
property of \emph{positivity}. Namely, if the function $a$ is nonnegative,
then for any $N,\,\sigma$, the operator $\OpAW(a)$ is positive.

These quantisations will be easy to compare once we have expressed
the anti-Wick quantisation in terms of the Weyl one \cite{BonDB}.
\begin{lem}\label{lem:antiWick}
Using the quadratic form $Q_\sigma(\bk):=\sigma\,k_1^2+\sigma^{-1}\,k^2_2$,
one has the following
expression for the anti-Wick quantisation:
\bequ\label{e:quant:tre}
\forall f\in L^1(\t2),\qquad
\OpAW(f)=\sum_{\bk\in\IZ^2}\tilde f(\bk)\;
\e^{-\frac{\pi}{2N}Q_\sigma(\bk)}\,T(\bk)\,.
\eequ
Equivalently, $\OpAW(f)=\OpW(f^\sharp)$, where the function $f^\sharp$ is obtained by
convolution of $f$ (on $\IR^2$) with the Gaussian kernel
\bequ\label{e:kernel1}
K_{N,\sigma}(\bx)\coloneq 2N\; \e^{-2\pi N Q_\sigma(\bx)}\,.
\eequ
\end{lem}
\dimostrazione
To prove this lemma, it is sufficient to
show that for any $\bk_0\in \IZ^2$, the anti-Wick
quantisation on $\hn$ of the Fourier mode $e_{\bk_0}(\bx)$ reads:
\bequ\label{e:antiWick}
\OpAW(e_{\bk_0})=\e^{-\frac{\pi}{2N}Q_\sigma(\bk_0)}\;T(\bk_0)\,.
\eequ
This formula has been proven in \cite[Lemma 2.3 ({\it ii})]{BonDB}, yet we
give here its proof for completeness. 
The idea is to decompose $\OpAW(e_{\bk_0})$ in the basis
$\{T(\bk),\ \bk\in\IZ_N^2\}$,
using the Hilbert-Schmidt scalar product \eqref{e:HS}. That is, we need to compute
\bequ\label{e:element}
\Big\la T(\bk), \OpAW(e_{\bk_0})\Big\ra
=\int_{\t2} e_{\bk_0}(\bx)\,\la \psi_{\bx,\sigma,\t2},T(\bk)^\dagger\,\psi_{\bx,\sigma,\t2}\ra\,\rmd\bx.
\eequ
The overlaps between torus coherent states derive from the overlaps between
plane coherent states,
which are simple Gaussian integrals:
$$
\forall \bx,\by\in\IR^2,\quad
\la \Psi_{\by,\sigma},\Psi_{\bx,\sigma}\ra_{\R^2}=
\e^{\i \frac{\by\wedge \bx}{2\hbar}}\,\la \Psi_{0,\sigma},\hat T_{\bx-\by}\,\Psi_{0,\sigma}\ra_{\R^2}=
\e^{\i \frac{\by\wedge \bx}{2\hbar}}\,\e^{-\frac{Q_\sigma(\bx-\by)}{4\hbar}}\,,
$$
Using the projector \eqref{e:projector}, we get
\begin{equation*}
\begin{split}
\la \psi_{\bx,\sigma,\t2}, \hat T_{\bk/N}\,\psi_{\bx,\sigma,\t2}\ra &=
\sum_{\bm\in\IZ^2} (-1)^{N m_1 m_2}\,\la \Psi_{\bx,\sigma},
\hat T_{\bk/N}\,\hat T_{\bm}\,\Psi_{\bx,\sigma}\ra_{\R^2} \\
&=\sum_{\bm\in\IZ^2} (-1)^{N m_1 m_2+\bm\wedge\bk}\,\e^{2\i\pi(\bx\wedge(\bk+N\bm))}\,
\e^{-\frac{\pi N}{2}Q_\sigma(\bm+\bk/N)}\,.
\end{split}
\end{equation*}
We insert this expression in the RHS of \eqref{e:element} (and remember that $N$ is even):
$$
\Big\la T(\bk), \OpAW(e_{\bk_0})\Big\ra
=\sum_{\bm\in\IZ^2}\delta_{\bk_0,\bk+N\bm}\,(-1)^{\bm\wedge\bk}\,
\e^{-\frac{\pi N}{2}Q_\sigma(\bm+\bk/N)}\,.
$$
This expression vanishes unless $\bk=\bk_1$, the unique
element of $\IZ^2_N$ s.t. $\bk_1=\bk_0+N\bm_1$ for
some $\bm_1\in\IZ^2$.
>From the orthonormality of the basis $\{T(\bk):\,\bk\in\IZ_N^2\}$, this shows that
$\OpAW(e_{\bk_0})=(-1)^{\bm_1\wedge\bk_1}\,\e^{-\frac{\pi N}{2}Q_\sigma(\bk_0)}\,T(\bk_1)
=\e^{-\frac{\pi N}{2}Q_\sigma(\bk_0)}\,T(\bk_0)$.
\qed

A simple property of these quantisations is the semi-classical behaviour of
the \emph{traces} of quantized observables:

\begin{lem}\label{lem:traces}
For any integer $M\geq 3$,
\bequ\label{e:traceWeyl}
\forall f\in C^\infty(\t2),\qquad \frac{1}{N}\,\Tr(\OpW(f))=\int_{\t2}f(\bx)\,\rmd\bx+
\cO_M\Big(\frac{\norm{f}_{C^M}}{N^M}\Big)\,.
\eequ
For the anti-Wick quantisation, we have:
\bequ\label{e:AWtrace}
\forall f\in L^1(\t2),\qquad
\frac{1}{N}\,\Tr(\OpAW(f))
=\int_{\t2}f(\bx)\,\rmd\bx+\cO(\norm{f}_{L^1}\,\e^{-\frac{\pi N}{2}\min(\sigma,1/\sigma)})\,.
\eequ
\end{lem}
\dimostrazione
The first identity uses
the fact that on the space $\hn$,
$$
\frac1N\,\Tr\, T(\bk)=\begin{cases}1& \text{if}\ \bk=N\bm\ \mbox{for some}\ \bm\in\IZ^2,\\
0&\mbox{otherwise.}\end{cases}
$$
The error term in \eqref{e:traceWeyl} is bounded above by
$\sum_{\bm\in\IZ^2\setminus\{0\}} |\tilde f(N\bm)|$.
Now, the Fourier coefficients of a smooth function satisfy
\bequ\label{e:Fourier-decay}
\forall M\geq 1,\quad\forall\bk\in\IZ^2,\qquad|\tilde{f}(\bk)|\ll_M
\frac{\norm{f}_{C^M}}{(1+|\bk|)^M}\,.
\eequ
Using this upper bound (with $M\geq 3$) in the above sum yields \eqref{e:traceWeyl}.

In the anti-Wick case,
each term $|\tilde f(N\bm)|\leq\norm{f}_{L^1}$ of the sum is multiplied by
$\e^{-\frac{\pi N}{2}Q_\sigma(\bm)}\leq \e^{-\frac{\pi N}{2}\min(\sigma,1/\sigma)|\bm|^2}$,
which yields \eqref{e:AWtrace}.
\qed

\bigskip

We will now compare the Weyl and anti-Wick quantisations in the operator norm. We give
two estimates, corresponding to the two classes of functions
described in the introduction of this section.

\begin{pro} \label{prop:weyl-AW}$ $

\noindent I) For any $f\in C^{\infty}(\T^2)$ and $\sigma>0$,
\bequ
\label{e:diff-estimate}
\norm{\OpW(f)-\OpAW(f)}\ll
\norm{f}_{C^5}\, \frac{\max\{\sigma,\sigma^{-1}\}}{N}\,.
\eequ
Here $\sigma$ may depend arbitrarily on $N$.

\noindent II) Let $\balpha\in \IR_+^2$, $|\balpha|<1$ and
assume that $\sigma>0$
may depend on $N$ such that the quantity
\bequ\label{e:MNsigma}
\hbar_{\balpha}(N,\sigma)\coloneq \max
\Big(\frac{N^{2\alpha_1-1}}{\sigma},N^{\alpha_1+\alpha_2-1},\sigma\,N^{2\alpha_2-1}\Big)
\eequ
goes to zero as $N\to\infty$.
Then there exists a seminorm $\cN_{\balpha}$ on the space $S_{\balpha}(\t2)$ such that,
for any $f=f(\cdot,N)\in S_{\balpha}(\t2)$, one has:
\bequ\label{e:diff2}
\forall N\geq 1,\qquad
\norm{\OpAW(f(\cdot,N))-\OpW(f(\cdot,N)}\ll\cN_{\alpha}(f)\,\hbar_{\balpha}(N,\sigma)\,.
\eequ
\end{pro}
\begin{remark}  The effective ``small parameter'' $\hbar_{\balpha}(N,\sigma)$
will be small as $N\to\infty$ only if
three conditions are simultaneously satisfied:-
\begin{itemize}
\item $|\balpha|=\alpha_1+\alpha_2<1$,
\item $N^{2\alpha_1}<<N \sigma$,
\item $N^{2\alpha_2}<<N/\sigma$.
\end{itemize}
These conditions mean that the horizontal and vertical widths of the kernel \eqref{e:kernel1}
must be small compared to the typical scale of fluctuations of $f$ in the respective directions.
The conditions $N\sigma \geq 1$, $N/\sigma\geq 1$ assumed in section~\ref{s:CS} are
therefore automatically satisfied.
\end{remark}


\noindent{\bf Proof of proposition~\ref{prop:weyl-AW}.}\phantom{X}

We start with the first (simple) part.
Our main ingredient is lemma~\ref{lem:antiWick}. By Taylor's theorem,
\bequ
\forall\bk\in\IZ^2,\qquad
\e^{-\frac{\pi}{2N}Q_\sigma(\bk)}= 1 + \cO\!\left(\frac{Q_\sigma(\bk)}{N}
\right)
= 1+ \cO\!\left(\frac{\max\{\sigma,\sigma^{-1}\}}{N} |\bk|^2\right)\,,
\label{e:quant:due}
\eequ
where the implied constant is independent of $\bk$.
Substituting \eqref{e:quant:due} into \eqref{e:quant:tre}, the first term
gives the Weyl quantisation of $f$. Using the bounds~\eqref{e:Fourier-decay}
with $M=5$, we obtain the first part of the proposition:
\bequ
\begin{split}
  \| \OpAW(f) - \OpW(f) \| &\ll \frac{\max\{\sigma,\sigma^{-1}\}}{N}
\sum_{\bk\in\IZ^2}
 |\tilde{f}(\bk)|\,|\bk|^2 \nonumber \\
&\ll  \frac{\max\{\sigma,\sigma^{-1}\}}{N}
\sum_{\bk\in\IZ^2}
 \frac{\|f\|_{C^5}}{(1+|\bk|)^5}\,|\bk|^2 \nonumber \\
&\ll \frac{\max\{\sigma,\sigma^{-1}\}}{N}\,\| f \|_{C^5} \,.
\end{split}
\eequ

\medskip

The second part of the proposition requires more care. We first need to control the
norm of the Weyl operator.
\begin{lem}\label{lem:CV}
Take any $\balpha,\,\bbeta\in\IR^2_+$ such that $|\bbeta|=1$ and $\bbeta\geq\balpha$ (i.e.
$\beta_i\geq\alpha_i$, $i=1,2$).
Then, for any function
$f=f(\cdot,N)\in S_{\balpha}(\t2)$, we have
\bequ\label{e:CV-alpha}
\norm{{\rm Op}^{{\rm W}}_{N}(f(\cdot,N))}\ll
\sum_{\gamma_1,\,\gamma_2=0}^1
C_{\balpha,\bga}(f)\;N^{-\bga\cdot(\bbeta-\balpha)}\,,
\eequ
and the implied constant is independent of $\balpha$,$\bbeta$.
\end{lem}
\dimostrazione
This lemma is a simple consequence of
the Calder\'on-Vaillancourt theorem, a sharp form of which was obtained in \cite{Boul}.
Assume $f$ is a smooth function on $\IR^2$ such that
$\partial^{\bga}f$ is uniformly bounded for all $\bga$ with $\gamma_1,\,\gamma_2\in\{0, 1\}$.
Then, its Weyl quantisation on $L^2(\IR)$ for $\hbar=1$ is a bounded operator, and:
\bequ\label{e:CV-R}
\norm{{\rm Op}^{{\rm W},\IR^2}_{\hbar=1}(f)}\leq C\,
\sum_{\gamma_1,\gamma_2=0}^1 \norm{\partial^{\bga}f}_{C^0(\IR^2)}\,.
\eequ
Here $\norm{\cdot}$ is the norm of bounded operators on $L^2(\IR)$, and $C$ is
independent of $f$.

Now, we use the scaling properties of the Weyl
quantisation\footnote{We thank N.~Anantharaman for pointing out
to us this scaling argument.}.
For any $\beta\in [0,1]$ and $\hbar>0$ we define
$$
f_{\hbar,\beta}(q,p):=f(\hbar^\beta q,\hbar^{1-\beta}p)\,.
$$
Then, if $U_{\hbar,\beta}$ is the dilation operator
$U_{\hbar,\beta} \psi(q) =\hbar^{\beta/2}\psi(\hbar^{\beta}q)$,
we have \cite[page 60]{Ma}
\begin{equation}
U_{\hbar,\beta}\, {\rm Op}^{{\rm W},\IR^2}_{\hbar}(f)\, U_{\hbar,\beta}^{-1}
={\rm Op}^{{\rm W},\IR^2}_{\hbar=1}(f_{\hbar,\beta})\,.
\end{equation}
Applying \eqref{e:CV-R} to $f_{\hbar,\beta}$, we obtain
$$
\forall\hbar>0,\qquad \norm{{\rm Op}^{{\rm W},\IR^2}_{\hbar}(f)}\leq
C\,\sum_{\gamma_1,\gamma_2=0}^1
\norm{\partial^{\bga}f}_{C^0(\IR^2)}\,\hbar^{\beta\gamma_1+(1-\beta)\gamma_2}\,.
$$
In the case $\hbar=(2\pi N)^{-1}$ we apply this bound to a function $f\in S_{\balpha}(\t2)$,
selecting
$\bbeta=(\beta,1-\beta)$ such that $\bbeta\geq\balpha$:
we then obtain the upper bound of \eqref{e:CV-alpha}
for $\OpWR(f)$.
The inequality \eqref{e:plane-torus} shows that
this bound applies as well to the Weyl operator on $\hn$.
\qed

\bigskip

Equipped with this lemma, we can now prove the second part of
proposition~\ref{prop:weyl-AW}.
>From the Taylor expansion
$$
|f(\bx+\by)-f(\bx)-(\by\cdot\nabla) f(\bx)|\leq
\frac{1}{2}\;\max_{0\leq t\leq 1}\left\{\big|(\by\cdot\nabla)^2 f(\bz)\big|\,,\,\bz=\bx+t\by\right\}\,
$$
and lemma~\ref{lem:antiWick},
one easily checks that for any $f\in C^\infty(\t2)$,
$$
\norm{f^\sharp-f}_{C^0}\leq \frac{1}{8\pi N}\Big(
\frac1\sigma\,\norm{\partial^2_q f}_{C^0}+2\norm{\partial_q\partial_p f}_{C^0}+
\sigma\,\norm{\partial^2_p f}_{C^0}\Big)\,.
$$
Since differentiation commutes
with convolution, one controls all derivatives:
\bequ\label{e:combination}
\forall\bga\in\IN^2_0,\qquad
\norm{\partial^{\bga}(f^\sharp-f)}_{C^0}\leq \frac{1}{8\pi N}\Big(
\frac1\sigma\,\norm{\partial^{\bga+(2,0)} f}_{C^0}+2\norm{\partial^{\bga+(1,1)} f}_{C^0}+
\sigma\,\norm{\partial^{\bga+(0,2)} f}_{C^0}\Big)\,.
\eequ
For $f=f(\cdot,N)\in S_{\balpha}(\t2)$,
this estimate implies:
\bal\label{e:semin-diff}
\norm{\partial^{\bga}(f^\sharp-f)}_{C^0}&\leq N^{\balpha\cdot\bga}\,
\Big( \frac{N^{2\alpha_1-1}}{\sigma}\,C_{\balpha,\bga+(2,0)}(f)\nonumber \\
&\qquad\qquad\qquad+ N^{\alpha_1+\alpha_2-1}\,C_{\balpha,\bga+(1,1)}(f)+
\sigma\,N^{2\alpha_2-1}\,C_{\balpha,\bga+(0,2)}(f)\Big)\nonumber\\
&\leq N^{\balpha\cdot\bga}\,\hbar_{\balpha}(N,\sigma)\,
\Big(C_{\balpha,\bga+(2,0)}(f)+ C_{\balpha,\bga+(1,1)}(f)
+C_{\balpha,\bga+(0,2)}(f)\Big)\,.
\end{align}
Here we used the parameter $\hbar_{\balpha}(N,\sigma)$ defined in \eqref{e:MNsigma}.
This shows that the function
$
f^{\sharp,{\rm rem}}(\cdot,N)\coloneq \frac{1}{\hbar_{\balpha}(N,\sigma)}\,
\big(f^\sharp(\cdot,N)-f(\cdot,N)\big)
$
is also an element of $S_{\balpha}(\t2)$, with seminorms dominated by
seminorms of $f$. Applying lemma~\ref{lem:CV} to that function and taking any
$\bbeta\geq\balpha$, $|\bbeta|=1$, we get
$$
\norm{\OpAW(f(\cdot,N))-\OpW(f(\cdot,N))}\ll
\hbar_{\balpha}(N,\sigma)\,
\sum_{|{\bga'}|\leq 2}\sum_{\gamma_1,\gamma_2=0}^1 \,C_{\balpha,\bga+\bga'}(f)\,.
$$
The seminorm stated in the theorem can therefore be defined as
\bequ\label{e:N_alpha}
\cN_{\balpha}(f)\coloneq
\sum_{|{\bga'}|\leq 2}\sum_{\gamma_1,\gamma_2=0}^1 \,C_{\balpha,\bga+\bga'}(f)\,.
\eequ
\qed


\subsection{Egorov estimates for the baker's map\label{s:Egorov}}$ $

We now turn to the proof of the Egorov property~\eqref{e:egor}.
Let us start with the case $n=1$. We assume that
$a$ is supported in the set $\cD_{1,\delta,\gamma}$ defined in
equation~(\ref{e:D1}), away from the discontinuity set $\cS_1$ of $B$.
\begin{pro} \label{prop:1-stepAW}
Let $0\lt \delta\lt 1/4$ and $0\lt \gamma \lt 1/2$. Assume that the support
of $a\in C^\infty(\t2)$ is contained in $\cD_{1,\delta,\gamma}$. Then, in the
classical limit,
\begin{equation*}
  \| \hat{B}_N\,\OpAW(a)\,\hat{B}_N^{-1} - {\rm Op}_N^{{\rm AW},\sigma/4}
   (a\circ B^{-1})\| \ll \norm{a}_{C^0}\, N^{5/4}\,\sigma^{1/4}\,\e^{-\pi N \theta}\,,
\end{equation*}
uniformly w.r.to $\delta$, $\gamma$, $\sigma\in[1/N,N]$. 
Here we took as before $\theta=\min(\sigma\del^2,\gamma^2/\sigma)$.
\end{pro}
\dimostrazione
For any normalised state $\phi\in\hn$, we consider the matrix element
\bequ
   \la\phi,\hat{B}_N \OpAW(a) \hat{B}_N^{-1} \phi\ra
=  N \int_{\t2} a(\bx)\, \la \phi ,\hat{B}_N\psi_{\bx,\sigma,\t2}
\ra \,\la\hat{B}_N \psi_{\bx,\sigma,\t2} , \phi\ra\, \rmd\bx\,.
\eequ
Using the quasi-covariance of coherent states localised in $\cD_{1,\delta,\gamma}$
(proposition~\ref{prop:zero}) and applying the Cauchy-Schwarz inequality, the RHS reads
\bequ
  N \int_{\t2} a(\bx)\, \langle \phi ,\psi_{B\bx,\sigma/4,\t2}
\ra\, \la \psi_{B\bx,\sigma/4,\t2} , \phi\ra\,\rmd\bx
+ \cO(\|a\|_{C^0} N^{5/4}\sigma^{1/4}\e^{-\pi N \theta})\,.
\eequ
The remainder is uniform with respect to the state $\phi$.
Through the variable substitution $\bx=B^{-1}(\by)$, this gives
\bequ
   \langle \phi, \hat{B}_N \OpAW(a) \hat{B}_N^{-1} \phi \ra
=  \langle \phi, {\rm Op}_N^{{\rm AW},\sigma/4}(a \circ {B}^{-1})\phi
\rangle + \cO(\norm{a}_{C^0} N^{5/4}\sigma^{1/4}\e^{-\pi N \theta}).
\eequ
Since the operators on both sides are self-adjoint, this identity implies the
norm estimate of the proposition.\qed

\begin{remark}
Here we used the property that the linear local dynamics is the same at each point
$\bx\in\t2\setminus\cS_1$
(expansion by a factor $2$ along the horizontal, contraction by $1/2$ along
the vertical). Were this not the case,
the state $\hat B_N \psi_{\bx,\sigma,\t2}$ would
be close to a coherent state at the point $B\bx$, but with a squeezing depending
on the point $\bx$. Integrating over $\bx$, we would get an anti-Wick quantisation of $a\circ B^{-1}$
with $\bx$-dependent squeezing, the analysis of which would be more complicated
(see \cite[Chap.~4]{Sch} for a discussion on such quantisations).
\end{remark}

\begin{figure}[htbp]
\begin{center}
\setlength{\unitlength}{5cm}
\begin{picture}(3,1.5)
\put(0,0.1){\includegraphics[angle=0,width=15.0cm,height=7cm]{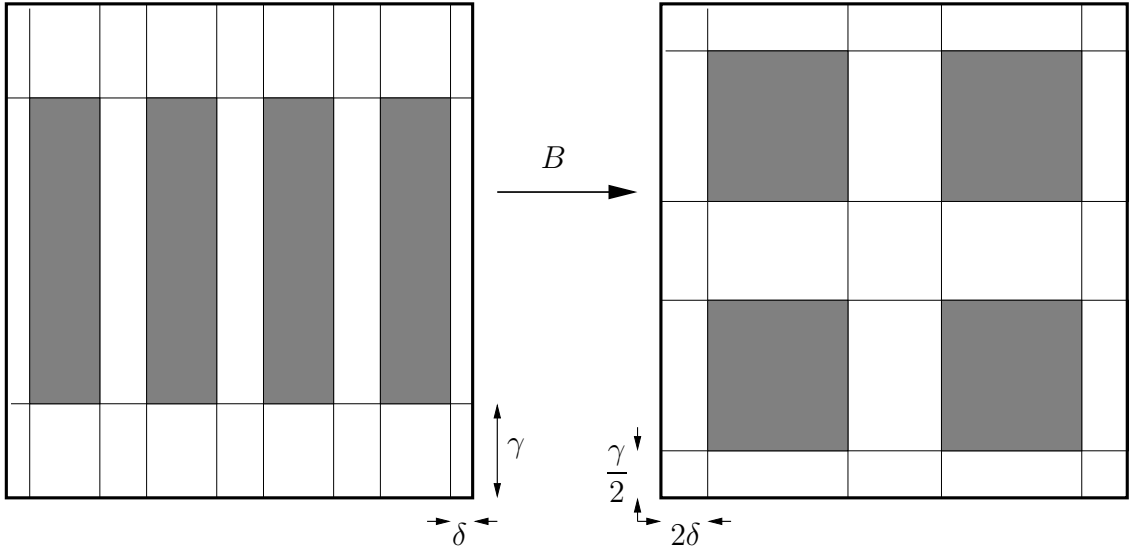}}
\put(1.43,1.05){$B$}
\put(1.195,0.05){$\delta$}
\put(1.34,0.295){$\gamma$}
\put(1.775,0.05){$2\delta$}
\put(1.6,0.22){$\displaystyle\frac{\gamma}2$}
\end{picture}
\caption{The action of the map $B$. On the left we show the set
$\cD_{2,\delta,\gamma}$ (shaded) and on the right is its image
under the action of $B$.}
\label{fig:zero}
\end{center}
\end{figure}

We now generalise to $n>1$. We assume that $a$ is supported
away from the set $\cS_n$ of discontinuities of $B^n$. More precisely,
for some $\delta\in(0,2^{-n-1})$ and $\gamma\in (0,1/2)$, we
define the following open set, generalizing \eqref{e:D1}:
$$
\cD_{n,\del,\gamma}\coloneq\left\{(q,p)\in\t2,\ \forall k\in\IZ\,,\
\Big|q-\frac{k}{2^n}\Big|>\del,\ p\in(\gamma,1-\gamma)\right\}\,.
$$
The evolution of the sets $\cD_{n,\delta,\gamma}$ through $B$
satisfies:
\begin{equation}
  \label{e:curlyD}
\forall j\in \{0,\ldots,n-1\},\qquad
B^j\cD_{n,\delta,\gamma}\subset \cD_{n-j,2^j\delta,\gamma/2^j}\,.
\end{equation}
This is illustrated for $n=2,\ j=1$ in figure \ref{fig:zero}.
If $a$ is supported in
$\cD_{n,\delta,\gamma}$, then the support of $a\circ B^{-j}$
is contained
in $\cD_{n-j,2^j\delta,\gamma/2^j} \subset \cD_{1,2^j\delta,\gamma/2^j}$.
So for each $0\leq j\lt n$, we can apply
proposition~\ref{prop:1-stepAW} to
the observable $a\circ B^{-j}$, replacing the parameters $\delta,\,\gamma,\, \sigma$
by their corresponding values at time $j$; we find that the parameter $\theta$ is
independent of $j$.
The triangle inequality then yields:
\begin{cor}
Let $n>0$ and for some $\delta\in(0,2^{-n-1})$, $\gamma\in(0,1/2)$,
let $a\in C^\infty(\t2)$ have support in $\cD_{n,\delta,\gamma}$.
Then, as $N\to\infty$,
\bequ\label{e:egorov-n-AW}
  \norm{\hat{B}_N^n\,\OpAW(a)\,\hat{B}_N^{-n} -
{\rm Op}_N^{{\rm AW},\sigma/4^n}(a\circ B^{-n})}
\ll \norm{a}_{C^0}\, N^{5/4}\,\sigma^{1/4}\e^{-\pi N \theta}\,.
\end{equation}
This estimate is uniform with respect to $n$, the parameters $\delta$, $\gamma$ in the
above ranges and and the squeezing
$\sigma\in [\frac{4^n}{N},N]$.
\end{cor}
\begin{remark}\label{r:time-window}
The requirement $N \theta>>1$, together with the allowed ranges for $\delta$, $\gamma$, 
impose the restriction  $\frac{4^n}{N}<<\sigma<<N$ $n\leq T_{\rm E}$. This
is possible only if $T_{\rm E}-n>>1$,
where $T_{\rm E}$ is the Ehrenfest time \eqref{e:Ehrenfest}.

We can reach times $n\sim T_{\rm E}(1-\eps)$ (with $\eps>0$ fixed) by taking the parameters
$\del=2^{-n-2}\asymp N^{-1+\eps}$, $\gamma\asymp 1$, $\sigma\asymp N^{1-\eps}$: in that
case, the argument of the exponential in
the RHS of equation~\eqref{e:egorov-n-AW} satisfies
$\pi N\theta \asymp N^{\eps}$, so that RHS decays
in the classical limit.
\end{remark}

We wish to obtain Egorov theorems where both terms correspond to a
quantisation with the same parameter $\sigma$, or the Weyl quantisation.
To do so, we will use
proposition~\ref{prop:weyl-AW} to replace the anti-Wick quantisations
by the Weyl quantisation. Using the first statement of that proposition,
we easily obtain
the following Egorov theorem:
\begin{thm}\label{thm:egorov-n}
Let $n>0$ and for some $\delta\in(0,2^{-n-1})$, $\gamma\in(0,1/2)$,
let $a\in C^\infty(\t2)$ have support in
$\cD_{n,\delta,\gamma}$. Then, in the limit $N\to\infty$, and for
any squeezing parameter $\sigma\in [\frac{4^n}{N},N]$, we have
\begin{multline}\label{e:egorov-n}
\norm{\hat{B}_N^n\,\OpW(a)\,\hat{B}_N^{-n} - \OpW(a\circ B^{-n})} \ll
\norm{a}_{C^0}\,N^{5/4}\,\sigma^{1/4}\,\e^{-\pi N \theta}\\
+\frac1N \Big(\max(\sigma,\sigma^{-1})\,\norm{a}_{C^5}
+\max\!\left(\frac{\sigma}{4^n},\frac{4^n}{\sigma}\right)\,\norm{a\circ B^{-n}}_{C^5}\Big).
\end{multline}
The implied constants are uniform in $n,\,\sigma,\,\del,\,\gamma$.
\end{thm}
If $n,\,\del,\,\gamma$ and the observable $a$ supported on $\cD_{n,\delta,\gamma}$
are independent of $N$, the RHS semi-classically
converges to zero if we simply take $\sigma=1$.
This is the ``finite-time'' Egorov theorem.

On the other hand, if we let $n$ grow with $N$, the function $a$ needs to change with $N$
as well (at least because its support needs to change).
In the next subsection we construct a specific family of functions
$\{a_n\}_{n\geq 1}$, each one supported away from $\cS_n$,
and compute the estimate \eqref{e:egorov-n} for this family.

\begin{remark}\label{r:past}
The same estimate holds if we replace $n$ by $-n$ on the LHS of \eqref{e:egorov-n}, and
replace $\sigma$ by $\sigma^{-1}$ on the RHS, including the definition of $\theta$.
Now, the function $a$ must be supported in the set
$\cD_{-n,\del,\gamma}$ obtained from $\cD_{n,\del,\gamma}$
by exchanging the roles of $q$ and $p$.

Indeed, using the unitarity of $\hat B_N$, we may interpret the
estimate \eqref{e:evol-0} as the quasi-covariant evolution of the
coherent state $\psi_{\by,\sigma',\t2}$ (where $\by=B\bx$, $\sigma'=\sigma/4$)
into the state $\psi_{B^{-1}\by,4\sigma',\t2}$, and the rest of the proof
identically follows.
\end{remark}


\subsection{Egorov estimates for truncated observables\label{s:opt}}

\subsubsection{A family of admissible functions}$ $

For future purposes (see the proof of theorem~\ref{main} in the next section), 
and in order to understand better the bound
\eqref{e:egorov-n}, we explicitly construct a sequence of functions $\{a_n\}_{n\geq 0}$,
each function being supported away from $\cS_n$. This sequence is
simply obtained by taking the products of
a fixed observable $a\in C^\infty(\t2)$ with cutoff functions
$\chi_{\del,n}$, which we now describe.

\begin{definition}\label{d:cutoff}
For some $0<\del<1/4$, we consider a
$\IZ$-periodic function $\tilde\chi_\del\in C^\infty(\IR)$ which vanishes for
$x\in[-\del,\del]\bmod\IZ$ and takes value $1$ for $x\in[2\del,1-2\del]\bmod\IZ$.

For any $n\geq 0$, we then define the following cutoff functions on $\t2$:
\begin{align*}
\chi_{\del,n}(\bx)&\coloneq\tilde\chi_\del(2^n\,q)\,\tilde\chi_\del(p)\,,\\
\chi_{\del,-n}(\bx)&\coloneq\tilde\chi_\del(2^n\,p)\,\tilde\chi_\del(q)\,.
\end{align*}
For any $n\in\IZ$, we split the observable $a\in C^\infty(\t2)$ into its ``good part''
$a_n(\bx)\coloneq a(\bx)\,\chi_{\del,n}(\bx)$ and its ``bad part''
$a_n^{\rm bad}(\bx)=a(\bx)\,(1-\chi_{\del,n}(\bx))$.
\end{definition}
One easily checks that $a_n$
is supported on $\cD_{n,\del/2^n,\del}$, while $a_n^{\rm bad}$ is supported
on a neighbourhood of $\cS_n$ of area $\cO(\del)$.

In light of remark \ref{r:past} we can, without loss of generality,
consider only times $n>0$.
For any multiindex $\bga\in\IN^2_0$, we have
\bequ\label{e:fluct1}
\norm{\partial^{\bga} a_n}_{C^0} \ll_{\bga}\norm{a}_{C^{|\bga|}}\, 2^{n\gamma_1}\,\delta^{-|{\bga}|}\,.
\eequ
When evolving $a_n$ through the map $B$, the derivatives grow along $p$ and decrease along $q$;
after $n$ iterations, $a_n\circ B^{-n}$ is still smooth, and
\bequ\label{e:fluct2}
\norm{\partial^{\bga} (a_n\circ B^{-n})}_{C^0}
\ll_{\bga}\norm{a}_{C^{|\bga|}}\, 2^{n\gamma_2}\,\delta^{-|{\bga}|}\,.
\eequ
These estimates show that the $C^5$-norms of $a_n$ and $a_n\circ B^{-n}$ (appearing on
the RHS of equation~\eqref{e:egorov-n}) are both of
order $2^{5n}/\del^5$. With our conventions, the parameter $\theta$ appearing in the RHS
of \eqref{e:egorov-n-AW} reads $\theta=\frac{\del^2}{\max(\sigma,4^n/\sigma)}$. We maximise
it by selecting $\sigma=2^n$.
With this choice, the upper bound \eqref{e:egorov-n} reads
\bequ\label{e:egorov-an}
\norm{\hat{B}_N^n\,\OpW(a_n)\,\hat{B}_N^{-n} - \OpW(a_n\circ B^{-n})} \ll
\norm{a}_{C^0}\, N^{5/4}\,2^{n/4}\,\e^{-\pi N \del^2/2^n}
+\frac{2^{6n}\,\norm{a}_{C^5}}{N\,\del^5}\,.
\eequ
Using remark~\ref{r:past}, the same estimate holds
if we replace $n$ by $-n$ on the LHS.

The last term of the RHS in \eqref{e:egorov-an} can semiclassically vanish only if
$|n|<\frac{T_{\rm E}}{6}$.
This time window, although not optimal (see the following subsection),
will be sufficient to prove theorem~\ref{main} in section~\ref{s:QE}.

Before that, in the last part of this section we will sharpen this estimate
by using the second part of proposition~\ref{prop:weyl-AW}: this will allow us
to prove a Egorov property up to times $|n|\leq (1-\epsilon)T_{\rm E}$, for any $\eps>0$.

\subsubsection{Optimised Egorov estimates}$ $

In this subsection we prove the following ``optimal'' Egorov theorem.

\begin{thm}\label{thm:egorov}
Choose $\eps>0$ arbitrarily small, and consider any observable $a\in C^\infty(\t2)$. For
any $N\geq 1$ and $n\in\IZ$,
construct the ``good part'' $a_n$ of that observable using definition~\ref{d:cutoff}
with a width $\del(N)\geq \min(N^{-\eps/4},1/10)$.

Then, the following Egorov estimate holds: there exists $C>0$ (independent
 of $a$, $\eps$) and
$N(\eps)>0$ such that for any $N\geq N(\eps)$ and any time $|n|\leq (1-\eps)T_{\rm E}$,
\bequ
\label{e:egorov-opt}
\norm{\hat B_N^{n}\,\OpW(a_n)\,\hat B_N^{-n}-\OpW(a_n\circ B^{-n})}
\leq C\Big(\norm{a}_{C^0}\, N^{3/2}\,\e^{-\pi N^{\eps/2}}
+\frac{\norm{a}_{C^{4}}}{N^{\eps/2}}\Big)\,.
\eequ
\end{thm}
\dimostrazione
We only treat the case $n\geq 0$, finally invoking the time-reversal symmetry
as in remark~\ref{r:past}.

We consider $\eps>0$ fixed, and define $N(\ep)$ through the equation $N(\ep)^{-\ep/4}=1/10$.
We then take $N\geq N(\ep)$
and consider any positive time $n\leq (1-\eps)T_{\rm E}$.

The improvement over theorem~\ref{thm:egorov-n}
will be a sharper bound for the norms
$\norm{\OpAW(a_n)-\OpW(a_n)}$ and
$\norm{{\rm Op}_N^{{\rm AW},\sigma/4^n}(a_n\circ B^{-n})-\OpW(a_n\circ B^{-n})}$.
Using the rescaled time $t=\frac{n}{T_{\rm E}}$ and
the property $\del(N)\geq N^{-\eps/4}$, the
bound \eqref{e:fluct1} on derivatives of $a_n$ reads:
$$
\norm{\partial^{\bga} a_n}_{C^0}
\ll_{\bga}\,\norm{a}_{C^{|\bga|}}\,2^{n\gamma_1}\,N^{\frac{\ep}{4}|\bga|}
=\norm{a}_{C^{|\bga|}}\,N^{t\gamma_1}\,N^{\frac{\ep}{4}|\bga|}\,.
$$
Thus, the derivatives of $a_n$ scale as those of an $N$-dependent function
in the space $S_{\balpha_t}(\t2)$, where $\balpha_t:=(t+\eps/4,\eps/4)$.
As in the former subsection, we must take $\sigma=2^n=N^t$ to
minimise the remainder. The second part of
proposition~\ref{prop:weyl-AW} applied to a function in $S_{\balpha_t}(\t2)$
yields a ``small parameter''
$\hbar_{\balpha_t}(N,2^n)=N^{t+\ep/2-1}$,
so that the difference between the two quantisations of $a_n$ is bounded as
$$
\norm{\OpW(a_n)-{\rm Op}_N^{{\rm AW},2^n}(a_n)}\ll
\norm{a}_{C^4}\,N^{t+\eps/2-1}\,.
$$
Similar considerations using \eqref{e:fluct2} show that
$\norm{\OpW(a\circ B^{-n})-{\rm Op}_N^{{\rm AW},2^{-n}}(a\circ B^{-n})}$
is bounded by the same quantity.
The argument of the exponential in equation~\eqref{e:egorov-n-AW}
takes the value
$N\theta=N\del^2/2^n\geq N^{1-t-\ep/2}$, so that the full estimate reads:
$$
\norm{\hat B_N^{n}\,\OpW(a_n)\,\hat B_N^{-n}-\OpW(a_n\circ B^{-n})}
\ll \norm{a}_{C^0}\, N^{3/2}\,\e^{-\pi N^{1-t-\eps/2}}
+\frac{\norm{a}_{C^{4}}}{N^{1-t-\eps/2}}\,.
$$
We obtain the bound \eqref{e:egorov-opt} uniform in $n$ by noticing that for the time window we
consider, $N^{1-t-\eps/2}\geq N^{\ep/2}$.
\qed

Our reason for believing that this estimate is ``optimal'' lies in
remark~\ref{r:time-window}:
we evolve states which stay away from the discontinuity set $\cS_1$
along their evolution. Since any state satisfies $\Delta q\Delta p\gtrsim
\frac12\hbar$ due to Heisenberg's uncertainty principle,
and $\Del q$ doubles at each time step, it is impossible
for such a state to remain away from $\cS_1$ during a time window larger than $T_{\rm E}$.

Besides, at the time $T_{\rm E}$ the ``good part'' $a_n$ oscillates on a
scale $\approx\hbar$ in the $q$ direction, so it behaves more like a Fourier integral
operator than an observable (pseudo-differential operator).


\section{Quantum Ergodicity\label{s:QE}}

For any even $N$, we denote by $\{\varphi_{N,j}\}$ the eigenvectors of $\hat B_N$ (if
some eigenvalues happen to be degenerate, which seems to be ruled out
by numerical
simulations, take an arbitrary orthonormal eigenbasis).
Let us consider a fixed real-valued observable $a\in C^\infty(\t2)$ satisfying
$\int_{\t2} a(\bx)\, \rmd\bx=0$.  Quantum
ergodicity follows if we prove that the quantum variance
\bequ
S_2(a,N)=\frac{1}{N}\sum_{j=1}^N |\la
\varphi_{N,j},\OpW(a)\,\varphi_{N,j}\ra|^2\Nto8 0.
\eequ

One method to prove this limit for our quantised baker's map would
be to apply the methods of \cite{MOK}: one only needs
the Egorov property (theorem~\ref{thm:egorov-n}) for finite times $n$,
and the classical \emph{ergodicity} of $B$.
However, this method seems unable to give information about the rate
of decay of the variance.

In order to prove the upper bound stated in theorem~\ref{main},
we will rather adapt the method used in \cite{Z2,Sch2} to our
discontinuous map.
This method requires the \emph{correlation functions}
of the classical map to decay sufficiently fast, which is the
case here (equation~\ref{e:mixing}).

\bigskip

\noindent{\bf Proof of Theorem \ref{main}.}\phantom{X}

To begin with, we consider the function
\begin{equation}
  \label{e:g}
  g(x)\coloneq 2\left ( \frac{1-\cos x}{x^2} \right)
\end{equation}
and its Fourier transform
$$
\hat{g}(k)=\int_{-\infty}^{\infty} g(x)\,\e^{-2\pi\i kx}\,\rmd x=
\begin{cases}
2\pi (1-|k|), & \mbox{for $-1\leq k\leq 1$,} \\
0, &\mbox{elsewhere.}
\end{cases}
$$
For any $T\geq 1$, we use it to construct the
following periodic function:
$$
f_T(\theta)\coloneq\sum_{m\in\Z} g(T(\theta + m)).
$$
$f_T$ admits the Fourier decomposition
$f_T(\theta)=\sum_{k\in\Z} \hat{f}_T(k)\, \e^{2\pi\i k \theta}$, where
$$
\hat{f}_T(k)=
\begin{cases}
\frac{2\pi}{T}\left(1-\frac{|k|}{T}\right) &\mbox{for $-T\leq k\leq T$,}\\
0 & \mbox{for $|k| >T$.}
\end{cases}
$$
Using this function, one may easily prove the following lemma \cite{Sch2}.
\begin{lem}\label{lem:roman}
With notations described above, for any even $N\geq 2$ and $T\geq 1$ one has
\begin{equation*}
S_2(a,N)\leq
\sum_{n\in\IZ} \hat f_T(n) \;\frac{1}{N}\,\Tr\Big(\OpW(a)\,
\hat B_N^n\, \OpW(a)\,\hat B_N^{-n}\Big)\,.
\end{equation*}
Notice that the terms in the sum on the RHS vanish for $|n|>T$.
\end{lem}
\dimostrazione
Let $\{\varphi_j\}$ be the eigenbasis of $\hat{B}_N$,
with $\hat{B}_N \varphi_j = \e^{2\pi\i \theta_j}\,\varphi_j $. Then
one has
\begin{equation*}
\Tr\Big(\OpW(a)\,\hat B_N^n\, \OpW(a)\,\hat B_N^{-n}\Big)
= \sum_{j,k=0}^{N-1} \e^{2\pi\i n (\theta_k-\theta_j)}\
|\langle\OpW(a)\varphi_j,\varphi_k\rangle|^2\,.
\end{equation*}
Multiplying by $\hat{f}_T(n)$ and summing over $n$, we get,
\begin{equation*}
\begin{split}
  \sum_{n\in\IZ}  \hat{f}_T(n)\;  \Tr\Big(\OpW(a)\,
\hat B_N^n\, \OpW(a)\,\hat B_N^{-n}\Big)
&= \sum_{j,k=0}^{N-1}
f_T(\theta_k-\theta_j)\
|\langle\OpW(a)\varphi_j,\varphi_k\rangle|^2 \\
&=\sum_{j=0}^{N-1} f_T(0)\
|\langle\OpW(a)\varphi_j,\varphi_j\rangle|^2 \\
&\qquad\qquad
+\sum_{j\neq k}  f_T(\theta_k-\theta_j)\
|\langle\OpW(a)\varphi_j,\varphi_k\rangle|^2 \\
&\geq N\;S_2(a,N)\,.
\end{split}\end{equation*}
The final inequality follows from the positivity of
$f_T$ and the property $f_T(0)\geq 1$. \qed

\bigskip

To prove the theorem we will estimate the
traces appearing in lemma~\ref{lem:roman}. Due to the support
properties of $\hat f_T$, only
the terms with $n\in [-T,T]$ will be needed. We take the time $T$ depending on $N$,
precisely as
$$
T=T(N)\coloneq\frac{T_{\rm E}}{11}\,,
$$
where $T_{\rm E}$ is the Ehrenfest time \eqref{e:Ehrenfest}.
For each $n\in\IZ\cap [-T,T]$, we will apply the Egorov
theorem~\ref{thm:egorov-n}. We first
decompose $a$ into a ``good'' part $a_n$ and ``bad'' part $a_n^{\rm bad}$,
as described in definition~\ref{d:cutoff}:
\begin{equation}
  \label{e:goodbad}
  a=a_n+a_n^{\rm bad}\,,\quad a_n\coloneq a.\chi_{\del,n}\,,
\end{equation}
We let the width $\delta>0$ depend on $N$ as
$\del\asymp (\log N)^{-1}$. Therefore, for any $n\in[-T,T]$ we will have
$\frac{2^{|n|}}{\del}\ll N^{1/10}$.
As a result, the
bounds \eqref{e:fluct1} for the derivatives of $a_n$ read:
\bequ\label{e:bound-an}
\forall n\in \IZ\cap [-T,T],\qquad\norm{\partial^{\bga} a_n}_{C^0} \ll_{\bga} \norm{a}_{C^{|\bga|}}\,
N^{\frac{|\bga|}{10}}\,.
\eequ
Furthermore, the same bounds
are satisfied by the derivatives of $a_n^{\rm bad}$ and $a_n\circ B^{-n}$.

\medskip
We decompose the traces of lemma~\ref{lem:roman}
according to the splitting \eqref{e:goodbad}:
\begin{align}
  \label{e:tracegb}
   \Tr\Big(\OpW(a)\,
\hat B_N^n\, \OpW(a)\,\hat B_N^{-n}\Big)&=
 \Tr\Big(\OpW(a)\,
\hat B_N^n\, \OpW(a_n)\,\hat B_N^{-n}\Big)\\ &\qquad\qquad+
 \Tr\Big(\OpW(a)\,
\hat B_N^n\, \OpW(a_n^{\rm bad})\,\hat B_N^{-n}\Big). \nonumber
\end{align}
The second term in the RHS will be controlled by replacing $\OpW(a_n^{\rm bad})$ by
its anti-Wick quantisation:
\begin{equation}
   \Tr\Big(\OpW(a)\,
\hat B_N^n\, \OpW(a_n^{\rm bad})\,\hat B_N^{-n}\Big)=
\Tr\Big(\OpW(a)\,
\hat B_N^n\, {\rm Op}_N^{{\rm AW}\!,1}(a_n^{\rm bad})\,\hat B_N^{-n}
+\cR_N(n)\Big)\,.
\end{equation}
The remainder $\cR_N(n)$ is dealt with using part $I$ of
proposition~\ref{prop:weyl-AW},
together with the bounds~\eqref{e:bound-an} applied to $a_n^{\rm bad}$:
\begin{align}
\|\cR_N(n)\| &\leq \norm{\OpW(a)}\,
\norm{\OpW(a_n^{\rm bad})-{\rm Op}_N^{{\rm AW}\!,1}(a_n^{\rm bad})}\nonumber\\
&\ll \norm{\OpW(a)}\ \frac{\norm{a_n^{\rm bad}}_{C^5}}{N}
\ll \norm{\OpW(a)}\ \frac{\norm{a}_{C^5}}{N^{1/2}}\,. \label{e:curlyR1}
\end{align}

In order to compute
$ \Tr\Big(\OpW(a)\,
\hat B_N^n\, {\rm Op}_N^{{\rm AW}\!,1}(a_n^{\rm bad})\,\hat B_N^{-n}\Big)$, we
split the function $a_n^{\rm bad}$ into its positive and negative parts,
$a_n^{\rm bad}=a_{n,+}^{\rm bad}-a_{n,-}^{\rm bad}$, where $a_{n,\pm}^{\rm bad}\geq 0$.
We then use the following (standard) linear algebra lemma to estimate the trace:
\begin{lem} \label{lem:ABtrace}
  Let $A$, $B$ be self-adjoint operators on $\cH_N$, and assume $B$ is positive.
Then
  \begin{equation}
    \label{e:ABtrace}
     |\Tr(AB)|\leq \norm{A}\, \Tr(B).
  \end{equation}
\end{lem}
Since the anti-Wick operator ${\rm Op}_N^{{\rm AW}\!,1}(a_{n,+}^{\rm bad})$ is positive,
this lemma yields:
\begin{align*}
  \Big|\Tr\Big(\OpW(a)\,\hat B_N^n\,
{\rm Op}_N^{{\rm AW}\!,1}(a_{n,+}^{\rm bad})\,\hat B_N^{-n}\Big)\Big|
 &\leq \norm{\OpW(a)} \; \Tr\big({\rm Op}_N^{{\rm AW}\!,1}(a_{n,+}^{\rm bad})\big)\,,
\end{align*}
and similarly by replacing $a_{n,+}^{\rm bad}$ by $a_{n,-}^{\rm bad}$.
By linearity and $a_{n,+}^{\rm bad}+a_{n,-}^{\rm bad}=|a_{n}^{\rm bad}|$, we get
$$
\Big|\Tr\Big(\OpW(a)\,\hat B_N^n\,
{\rm Op}_N^{{\rm AW}\!,1}(a_{n}^{\rm bad})\,\hat B_N^{-n}\Big)\Big|
\leq \norm{\OpW(a)} \; \Tr\big({\rm Op}_N^{{\rm AW}\!,1}(|a_{n}^{\rm bad}|)\big)
$$
{}From equation~\eqref{e:AWtrace}, the trace on the RHS is equal to
$N\cdot\norm{a_{n}^{\rm bad}}_{L_1(\t2)}\big(1+\cO(\e^{-\pi
N/2})\big)$. Since $a_{n}^{\rm bad}$ is supported on a neighbourhood
of $\cS_n$ of area $\cO(\del)$, its $L^1$ norm is of order
$\cO(\del\,\norm{a}_{C^0})$. Using the Calder\'on-Vaillancourt
estimate $\norm{\OpW(a)}\leq C\,\norm{a}_{C^2}$, we have thus proven
the following bound for the second term in \eqref{e:tracegb}: \bequ
\frac1N  \Tr\Big(\OpW(a)\,\hat B_N^n\,\OpW(a_n^{\rm bad})\,\hat
B_N^{-n}\Big) \ll \norm{a}_{C^2}\left( \del\,\norm{a}_{C^0} +
\frac{\norm{a}_{C^{5}}}{N^{1/2}}\right)\,. \eequ

\bigskip

We now estimate the first term in \eqref{e:tracegb}. We write
\begin{equation}
   \Tr\Big(\OpW(a)\,
\hat B_N^n \OpW(a_n)\hat B_N^{-n}\Big)=
\Tr\Big(\OpW(a)\OpW(a_n\circ B^{-n})+\cR'_N(n)\Big)\,,
\end{equation}
and control the remainder $\cR^{\prime}_N(n)$ with the Egorov estimate \eqref{e:egorov-an},
remembering that $n\leq T_{\rm E}/11$:
\begin{align}
 \norm{\cR^{\prime}_N(n)} &\ll \norm{\OpW(a)}\,
\Big(\norm{a}_{C^0}\, N^{5/4}\,2^{n/4}\,\e^{-\pi N \del^2/2^n}
+\frac{2^{6n}\,\norm{a}_{C^5}}{N\,\del^5}\Big)\,\nonumber\\
&\ll \frac{\norm{a}^2_{C^{5}}}{N^{2/5}}\,.
\label{e:curlyR2}
\end{align}

The following lemma (proved in \cite[lemma~3.1]{MOK})
will allow us to replace the quantum product by a classical one.
\begin{lem}\label{lem:product}
There exists $C>0$ such that, for any pair $a,b \in C^{\infty}(\t2)$,
  \begin{equation}
    \label{e:product}
\forall N\geq 1,\qquad    \norm{\OpW(a)\,\OpW(b)-\OpW(ab)}
    \leq C\, \frac{\norm{a}_{C^4}\,\norm{b}_{C^4}}{N} \,.
  \end{equation}
\end{lem}
Using this lemma and the bounds~\eqref{e:bound-an}, we get
\begin{align}
  \Tr\Big(\OpW(a)\,\OpW(a_n\circ B^{-n})\Big)&=\Tr\Big(\OpW\big(a(a_n\circ B^{-n})\big)
+\cR^{\prime\prime}_N(n)\Big)\,,\nonumber\\
\text{with}\qquad
\norm{\cR^{\prime\prime}_N(n)} &\ll \frac{\norm{a_n}_{C^4}\;\norm{a_n\circ B^{-n}}_{C^4}}{N}
\ll \frac{\norm{a}^2_{C^{4}}}{N^{1/5}}\,.\label{e:curlyR3}
\end{align}
To finally estimate the trace of $\OpW\big(a(a_n\circ B^{-n})\big)$, we use
equation~\eqref{e:traceWeyl}
together with the estimates~\eqref{e:bound-an}:
$$
\frac1N \Tr\!\left(\OpW\big(a(a_n\circ B^{-n})\big)\right)
= \int_{\t2}a(a_n\circ B^{-n})(\bx)\,\rmd\bx
+\cO\Big(\frac{\norm{a}^2_{C^3}}{N^2}\Big)\,.
$$
It remains to compute the integral on the RHS.
We split it in two integrals, according to $a_n=a-a_n^{\rm bad}$.
The second integral can be bounded by
\begin{equation}
 \left| \int_{\t2} a(\bx)\,a_n^{\rm bad}( B^{-n}\bx)\,\rmd \bx\right|
\leq \norm{a}_{C^0}\,\norm{a_n^{\rm bad}}_{L^1} \ll \norm{a}^2_{C^0}\;\delta\,,
\end{equation}
while the first one reads
\begin{equation}
  \int_{\t2} a(\bx)\,a( B^{-n}\bx)\,\rmd \bx = {\mathcal K}_{a\,a}(n)\,.
\end{equation}
This integral is the classical autocorrelation function for the observable $a(\bx)$,
a purely classical quantity. At this point we must use the dynamical properties
of the classical baker's map $B$, namely its fast mixing properties
(see the end of section~\ref{s:classical}): for some $\Gamma>0$, the
autocorrelation decays (when $n\to\infty$) as
$$
{\mathcal K}_{a\,a}(n)\ll \norm{a}^2_{C^1}\, \e^{-\Gamma|n|}\,.
$$

Collecting all terms and using the properties of the function $\hat f_T$, 
the lemma~\ref{lem:roman}
finally yields the following upper bound:
\begin{align*}
S_2(a,N)&\ll  \norm{a}^2_{C^{5}}\, \sum_{n\in [-T,T]}|\hat f_T(n)|\
\Big( \e^{-\Gamma|n|} + \delta + \frac{1}{N^{1/5}}\Big)\\
&\ll \norm{a}^2_{C^{5}}\, \Big(\frac{1}{T}+\delta\Big)\,.
\end{align*}
Since we took $T\asymp \log N$ and $\del\asymp (\log N)^{-1}$,
this concludes the proof of theorem \ref{main}. \qed

\bigskip
\noindent{\bf Proof of corollary \ref{maincor}.}\phantom{X}
We start by picking an observable $a\in C^\infty(\t2)$, assuming
$\int a(\bx)\rmd\bx=0$.
For any decreasing sequence $\alpha(N)\Nto8 0$, 
Chebychev's inequality yields an upper bound 
on the number of eigenvectors
of $\hat B_N$ for which $|\la\varphi_{N,j},\OpW(a)\varphi_{N,j}\ra|> \alpha(N)$:
\begin{align}
  \frac{\#\big\{j\in\{1,\ldots,N\}\,:\,|\la \varphi_{N,j}, \OpW(a)\,\varphi_{N,j} \ra| >
 \alpha(N) \big\}}{N} \leq \frac{S_2(a,N)}{\alpha(N)^2}\,.
\end{align}
From the theorem \ref{main}, if we take
$\alpha(N)>> (\log N)^{-1/2}$, the above fraction converges to
zero. Defining $J_N(a)$ as the complement of the set 
in the above numerator, we obtain 
a sequence of subsets $J_N(a)\subset\{1,\ldots,N\}$ satisfying $\frac{\#J_N(a)}{N}\to 1$, 
such that the eigenstates
$\varphi_{N,j_N}$ with $j_N\in J_N(a)$ satisfy \eqref{e:QE}. 

Using a standard
diagonal argument \cite{CdV,HMR,Zel}, one can then extract subsets 
$J_N\subset\{1,\ldots,N\}$ independent
of the observable $a\in C^\infty(\t2)$, with $\frac{\#J_N}{N}\to 1$,
such that \eqref{e:QE} is satisfied for any $a\in C^\infty(\t2)$ if one takes $j_N\in J_N$.
\qed



\begin{thebibliography}{999999}

\bibitem[ALP{\.{Z}}]{Zycz}{R.~Alicki, A.~Lozinski, P.~Pakonski and K.~\.{Z}yczkowski (2004)
``Quantum dynamical entropy and decoherence rate'' {\it J.\ Phys.\ A} {\bf 37} 5157--5172.}

\bibitem[AA]{AA}{V.I.~Arnold and A.~Avez (1967) ``Problèmes ergodiques de la mécanique classique'',
Gauthier-Villars (Paris).}

\bibitem[BSS]{BSS}{A.~B\"acker, R.~Schubert and P.~Stifter (1998)
``Rate of quantum ergodicity in Euclidean billiards'', {\it Phys. Rev. E} {\bf 57} 5425--5447;
Erratum, {\it Phys.\ Rev.\ E} {\bf 58} 5192.}

\bibitem[BGP]{BGP}{D.~Bambusi, S.~Graffi and T.~Paul (1999)
``Long time semiclassical approximation of quantum flows:
a proof of the Ehrenfest time'' {\it Asymptot.\ Anal.} {\bf 21} 149--160.}

\bibitem[Bar]{Bar}{A.\ Barnett (2004)
``Asymptotic rate of quantum ergodicity in chaotic Euclidean billiards''
submitted to {\it Comm.\ Pure Appl.\ Math.}}

\bibitem[BdMOdA]{BasOz}{M.\ Basilio de Matos and A.\ M.\ Ozorio de Almeida (1995)
``Quantization of Anosov maps'' {\it Ann.\ Phys.} {\bf 237} 46--65.}

\bibitem[BonDB]{BonDB}
{F.~Bonechi and S.~De Bi\`evre  (2000)
``Exponential mixing and $\ln \hbar$ timescales in quantized hyperbolic maps on the torus'', 
{\it Commun. Math. Phys.} {\bf 211} 659--686.}

\bibitem[Boul]{Boul}{A.~Boulkhemair (1999)
``$L^2$ estimates for Weyl quantization'' {\it J.\ Funct.\ Anal.} {\bf 165} 173--204.}

\bibitem[BDB]{BDeB}{A.~Bouzouina  and S.~De~Bi\`{e}vre (1996)
``Equipartition of the eigenfunctions of quantized
ergodic maps on the torus'' {\it Commun. Math. Phys.} {\bf 178} 83--105.}


\bibitem[BR]{BR}{A.~Bouzouina and D.~Robert (2002)
``Uniform semi-classical estimates for the propagation of
quantum observables'' {\it Duke Math.\ J.} {\bf 111} 223--252.}


\bibitem[BV]{BV}{N.\ L.\ Balazs and A.\ Voros (1989) ``The quantized baker's transformation''
{\it Ann.\ Phys.} {\bf 190} 1--31.}

\bibitem[Ch]{Chernov}{N.I.~Chernov (1992) ``Ergodic and statistical properties of
piecewise linear hyperbolic automorphisms of the 2-torus'',
{\it J.\ Stat.\ Phys.} {\bf 69}, 111--134.}

\bibitem[CdV]{CdV}{Y.\ Colin~de~Verdi\'ere (1985) ``Ergodicit\'e et fonctions propres du
Laplacien'' {\it Commun.\ Math.\ Phys.} {\bf 102}  497--502.}

\bibitem[DBDE]{DBDE}{S.\ De~Bi\`{e}vre and M.\ Degli~Esposti (1998) ``Egorov
theorems and equidistribution of eigenfunctions
for the quantized sawtooth and Baker maps'' {\it Annales de l'Institut H.
Poincar\`{e}, Phys. Theor.} {\bf 69} 1--30.}

\bibitem[DEG]{DEG}{M.\ Degli~Esposti and S.\ Graffi (2003) ``Mathematical
aspects of quantum maps'' in M.\ Degli~Esposti and S.\ Graffi, editors
{\it The mathematical aspects of quantum maps}, volume 618 of
Lecture Notes in Physics, Springer, 2003, pp. 49--90.}

\bibitem[DEGI]{DEGI}{M.\ Degli~Esposti, S.\ Graffi and S.\ Isola (1995)
``Classical limit of the quantized hyperbolic toral
automorphism'' {\it Commun.\ Math.\ Phys.} {\bf 167}
471--507.}

\bibitem[DE${}^{+}$]{DEO'KW} {M.\ Degli~Esposti, S.\
O'Keefe and B.\ Winn (2005) ``A semi-classical study of the
Casati-Prosen triangle map'' {\it Nonlinearity} {\bf 18} 1073--1094.}

\bibitem[DS]{DS}M.~Dimassi and J.~Sj\"ostrand, {\it Spectral Asymptotics in the
Semi-Classical Limit}, Cambridge University Press, 1999.

\bibitem[EFK${}^{+}$]{EFK}{B.\ Eckhardt, S.\ Fishman, J.\ Keating, O.\ Agam,
J.\ Main, K.\ M\"uller (1995) ``Approach to ergodicity in
quantum wave functions'' {\it Phys.\ Rev.\  E} {\bf 52} (1995) 5893--5903.}

\bibitem[Fa]{Fa}{M.~Farris (1981) ``Egorov's theorem on a manifold with diffractive
boundary'' {\it Commun.\ Partial Differential Equations} {\bf 6} 651--687.}

\bibitem[FP]{FP} {M.\ Feingold and A.\ Peres (1986) ``Distribution of matrix
elements of chaotic systems'' {\it Phys.\ Rev.\ A} {\bf 34} 591--595.}

\bibitem[FDBN]{FDBN} {F.\ Faure, S.\ Nonnenmacher and S.\ De~Bi\`evre (2003)
``Scarred eigenstates for quantum cat maps of minimal periods''
{\it Commun.\ Math.\ Phys.} {\bf 239} 449--492.}

\bibitem[Fo]{Fo} {G.\ B.\ Folland {\it Harmonic analysis in phase space},
The Annals of mathematics studies {\bf 122}, Princeton University Press, 1989.}

\bibitem[GL]{GL}{P.~G\'erard and \'E.~Leichtnam (1993)
``Ergodic properties of eigenfunctions for the Dirichlet problem''
{\it Duke Math.\ J.} {\bf 71} 559--607.}

\bibitem[HB]{HB}{J.\ H.\ Hannay and M.\ V.\ Berry (1980) ``Quantisation of
linear maps on the torus---Fresnel diffraction by a periodic
grating'' {\it Physica D} {\bf 1} 267--290.}

\bibitem[Has]{Has}{H.H.~Hasegawa and W.C.~Saphir (1992) ``Unitarity and irreversibility
in chaotic systems'', {\it Phys.\ Rev.\ A} {\bf 46} 7401--7423.}

\bibitem[HMR]{HMR}{B.~Helffer, A.~Martinez and D.~Robert (1987) 
``Ergodicit\'e et limite semi-classique'' {\it Commun.\ Math.\ Phys.} {\bf 109}
313--326.}

\bibitem[Kap]{Kap}{L.~Kaplan and E.J.~Heller (1998) ``Linear and nonlinear theory
of eigenfunction scars'' {\it Ann.\ Phys.\ (NY)} {\bf 264} 171--206.}

\bibitem[KM]{KM}{J.P.~Keating and  F.~Mezzadri (2000) ``Pseudo-symmetries of
Anosov maps and spectral statistics''  {\it Nonlinearity} {\bf 13} 747--775.}

\bibitem[KR1]{KR1}{P.~Kurlberg and Z.~Rudnick (2001)
``Hecke theory and equidistribution for the quantization of
linear maps of the torus'' {\it Duke Math.\ J.} {\bf 103} 47--77.}

\bibitem[KR2]{KR2}{P.~Kurlberg and Z.~Rudnick (2001)
``On quantum ergodicity for linear maps of the torus''
{\it Commun.\ Math.\ Phys.} {\bf 222} 201--227.}

\bibitem[KR3]{KR3}{P.~Kurlberg and Z.~Rudnick (2005)
``On the distribution of matrix elements for the quantum cat map''
{\it Ann.\ Math.} {\bf 161} 489--507.}

\bibitem[Lak]{Lak}{A.~Lakshminarayan (1995)
``On the quantum baker's map and its unusual traces''
{\it Ann.\ Phys.\ (NY)} {\bf 239} 272--295.}

\bibitem[LV]{LebVor}{P.~Leb{\oe}uf and A.~Voros  (1990)
``Chaos revealing multiplicative representation of quantum eigenstates''
{\it J.\ Phys.\ A} {\bf 23} 1765--1774.}

\bibitem[Lin]{Lin} {E.\ Lindenstrauss (2004)
``Invariant measures and arithmetic quantum unique ergodicity''
To appear in {\it Ann.\ Math.}}

\bibitem[LS]{LS}{W.\ Luo and P.\ Sarnak
``Quantum variance for Hecke eigenforms''(2004)
{\it Ann. Sci. Ecole Norm. Sup.} {\bf 37} 769--799}

\bibitem[MO'K]{MOK}{J.\ Marklof and S.\ O'Keefe (2005)
``Weyl's law and quantum ergodicity for maps with divided phase
space''; appendix by S.\ Zelditch ``Converse quantum ergodicity''
{\it Nonlinearity} {\bf 18} 277--304.}

\bibitem[MR]{MR}{ J.~Marklof and Z.~Rudnick (2000) ``Quantum unique
ergodicity for parabolic maps'' {\it Geom.\ Func.\ Anal.} {\bf 10} 1554--1578.}

\bibitem[Ma]{Ma}{A.~ Martinez {\it An introduction to semiclassical and
microlocal analysis} Springer-Verlag, 2002.}

\bibitem[O'CTH]{O'CTH}{P.W.~O'Connor, S.~Tomsovic and E.J.~Heller (1992)
``Accuracy of semiclassical dynamics in the presence of chaos''
{\it J.\ Stat.\ Phys.} {\bf 68} 131--152.}

\bibitem[Per]{perelomov}{A.M.~Perelomov (1986)
``Generalized coherent states and their applications''
Springer Verlag (Heidelberg).}

\bibitem[Rob]{Rob}{D.~Robert (2004) ``Remarks on time dependent Schr\"odinger
equation, bound states and coherent states'', in 
{\it Multiscale methods in quantum mechanics}, 139--158, Trends Maths, Birkh\"auser
(Boston).}

\bibitem[Ros]{Rosen}{L.~Rosenzweig (2004)
{\it Quantum unique ergodicity for maps on $\t2$.} M.Sc.\ Thesis, Tel Aviv
University.}

\bibitem[RubSal]{RubSal}{R.~Rubin and N.~Salwen (1998)
``A Canonical Quantization of the Baker's Map'',
{\it Ann.\ Phys.\ (NY)} {\bf 269}  159--181.}

\bibitem[RudSar]{RS}{Z.~Rudnick and P.~Sarnak (1994) ``The behaviour of
eigenstates of arithmetic hyperbolic manifolds'' {\it Commun.\ Math.\ Phys.}
{\bf 161} 195--213.}

\bibitem[RuSo]{RuSo}{Z.~Rudnick and K.~Soundararajan, in preparation (2004)}

\bibitem[Sa]{Sa}{M.~Saraceno (1990)
``Classical structures in the quantized baker transformation''
{\it Ann.\ Phys.\ (NY)} {\bf 199} 37--60.}

\bibitem[SaVo]{SV1}{M.~Saraceno and A.~Voros (1994)
``Towards a semiclassical theory of the quantum baker's map''
{\it Physica D} {\bf 79} 206--268.}

\bibitem[Sar1]{Sar}{P.~Sarnak (2003), ``Spectra of Hyperbolic Surfaces''
{\it Bull.\ Amer.\ Math.\ Soc.} {\bf 40}, no.4, 441-478.}

\bibitem[Sar2]{sar2}{P.\ Sarnak {\it Quantum vesus classical fluctuations on
the modular surface}. Talk given at the meeting: ``Random Matrix Theory and
Arithmetic Aspects of Quantum Chaos'' at the Isaac Newton Institute,
Cambridge, June 2004.\\ Audio file available at
{\tt http://www.newton.cam.ac.uk/webseminars/}}

\bibitem[Schn]{Scn}{A.\ I.\ Schnirelmann (1974) ``Ergodic properties of
eigenfunctions'' {\it  Uspekhi Mat.\ Nauk.} {\bf 29}  181--182.}

\bibitem[Schu1]{Sch}{R.~Schubert (2001) {\it Semiclassical localization in phase space.}
Ph.D. Thesis, Universit\"{a}t Ulm.
Available at {\tt http://vts.uni-ulm.de}}

\bibitem[Schu2]{Sch2}{R.~Schubert ``Upper bounds on the rate of quantum ergodicity.''
Preprint 2005, available at {\tt arXiv:math-ph/0503045}.}

\bibitem[Wil]{Wil}{M.~Wilkinson (1987)
``A semiclassical sum rule for matrix elements of
classically chaotic systems'' {\it J.\ Phys.\ A} {\bf 9} 2415--2423.}

\bibitem[Zel1]{Zel}{S.~Zelditch (1987)
``Uniform distribution of eigenfunctions on compact hyperbolic
surfaces'' {\it Duke Math.\ J.} {\bf 55} 919--941.}

\bibitem[Zel2]{Z2}{S.~Zelditch (1994)
``On the rate of quantum ergodicity. I. Upper bounds'' {\it Commun.
Math. Phys.} {\bf 160} no. 1, 81--92.}

\bibitem[ZZ]{ZZ}{S.~Zelditch and M.~Zworski (1996)
``Ergodicity of eigenfunctions for ergodic billiards'' {\it Commun.\
Math.\ Phys.} {\bf 175} 673--682.}

\end{thebibliography}
\end{document}